\documentclass[twocolumn,tighten]{aastex62}

\newcommand{\rrab}{RR{\sl ab}}
\newcommand{\rrc}{RR{\sl c}}

\accepted{\today}

\submitjournal{AJ}

\shorttitle{Pulsating Variable Stars in Sextans}
\shortauthors{Vivas et al.}

\begin{document}

\title{The Population of Pulsating Variable Stars in the Sextans Dwarf Spheroidal Galaxy}

\correspondingauthor{A. K. Vivas}
\email{kvivas@ctio.noao.edu}

\author[0000-0003-4341-6172]{A. Katherina Vivas}
\affiliation{Cerro Tololo Inter-American Observatory, National Optical Astronomy Observatories, Casilla 603,
La Serena, Chile}

\author[0000-0003-3496-3772]{Javier Alonso-Garc{\'\i}a}
\affiliation{Centro de Astronom\'{i}a (CITEVA), Universidad de Antofagasta, Av. Angamos 601, Antofagasta, Chile}
\affiliation{Instituto Milenio de Astrof\'{i}sica, Santiago, Chile}

\author{Mario Mateo}
\affiliation{Department of Astronomy, University of Michigan, Ann Arbor, MI 48109, USA}

\author[0000-0002-7123-8943]{Alistair Walker}
\affiliation{Cerro Tololo Inter-American Observatory, National Optical Astronomy Observatories, Casilla 603,
La Serena, Chile}

\author{Brittany Howard}
\affiliation{Department of Physics and Astronomy, University of Victoria, PO Box 1700 STN CSC, Victoria, BC  V8W 2Y2, Canada}

\begin{abstract}

A large extension of the Sextans dwarf spheroidal galaxy, 7 sq degrees, has been surveyed for variable stars using the Dark Energy Camera at the Blanco Telescope in Cerro Tololo Inter-American Observatory, Chile. We report 7 Anomalous Cepheids, 199 RR Lyrae stars and 16 dwarf Cepheids in the field. This is only the fifth extra-galactic systems in which dwarf Cepheids have been systematically searched. Henceforth, the new stars increase the census of stars coming from different environments that can be used to asses the advantages and limitations of using dwarf Cepheids as standard candles in populations for which the metallicity is not necessarily known. The dwarf Cepheids found in Sextans have a mean period of 0.066 days, and a mean $g$ amplitude of 0.87 mags. They are located below the horizontal branch spanning a range of 0.8 mag, between $21.9 < g < 22.7$. The number of dwarf Cepheids in Sextans is low compared with other galaxies such as Carina, which have a strong intermediate-age population. On the other hand, the number and ratio of RR Lyrae stars to dwarf Cepheids is quite similar to Sculptor, a galaxy which, as Sextans, is dominated by an old stellar population. The dwarf Cepheid stars found in Sextans follow a well constrained Period-Luminosity relationship with an rms=0.05 mag in the $g$ band, which was set up by anchoring to the distance modulus given by the RR Lyrae stars. Although the majority of the variable stars in Sextans are located toward the center of the galaxy, we have found 2 RR Lyrae stars and 1 Anomalous Cepheid in the outskirts of the galaxy, which may be extra-tidal stars and suggest this galaxy may be undergoing tidal destruction. These possible extra-tidal variable stars share the same proper motions as Sextans, as seen by recent Gaia measurements. Two additional stars which we initially classified as foreground RR Lyrae stars may actually be other examples of Sextans extra-tidal Anomalous Cepheids, although radial velocities are needed to prove that scenario.

\end{abstract}


\section{Introduction} \label{sec:intro}

Below the horizontal branch (HB), an interesting group of pulsating variable stars can be found. These stars, named $\delta$ Scuti  (Sct) if metal-rich or SX Phoenicis (Phe) if metal-poor, have an important property: they follow a Period-Luminosity (PL) relationship \citep[][and reference therein]{sandage06,mcnamara11} and thus they could be used, in principle, as standard candles with precision similar to RR Lyrae stars ($\sim 5-7\%$). These stars can also be numerous. The Large Magellanic Clouds (LMC), for example, contains a couple of thousand of them \citep{garg10,poleski10}, while the Carina dwarf Spheroidal (dSph) galaxy have $\sim 400$, about 5 times the number of RR Lyrae stars in that galaxy \citep{vivas13,coppola15}. In the LSST era, these pulsating stars will be found to large distances and motivates further study of their properties to understand their use and limitations as extra-galactic standard candles and as tracers of structure within the halo of the Milky Way. 

The naming of the $\delta$ Sct and SX Phe type of variables stars has not been without controversy in the past \citep[see][for an extensive discussion]{catelan15}. In this paper we will use the term {\sl Dwarf Cepheid} (DC) to refer collectively to the High Amplitude $\delta$ Sct stars, or HADS, and the SX Phe variables \citep[following other works such as][]{mateo93,vivas13}. The reasoning behind the use of this nomenclature is that both stars occupy the same region in the Hertzsprung-Russell (H-R) diagram and, because dSph galaxies can contain a mix of stellar populations, it is not easy to know if, in these systems, those variable stars belong to a Population I (hence, if they are $\delta$ Sct stars), to a Population II (hence, if they are SX Phe stars) or to a mix of both. Furthermore, both type of stars share similar pulsational characteristics \citep{balona12}, and, particularly important in the context of this work, they seem to share the same PL relationship \citep{cohen12}. On the other hand, the evolutionary status of HADS and SX Phe stars could not be more different. HADS are thought to be main sequence stars that lie within the instability strip, while SX Phe stars are blue stragglers from old populations and have reached that position in the H-R diagram through binary evolution.

\citet{cohen12} suggested that the PL relationship of DC does not depend on color or metallicity. Their PL relationship was constructed by joining DC stars from galactic globular clusters and a few extra-galactic systems (LMC, Fornax and a small sample from Carina known at the time of that work); that is, it contains both HADS and SX Phe stars. The dispersion in the resulting PL relation is relatively large, $\sim 0.1$ mag, although the authors claim that part of the scatter may be due to calibration differences between the multiple data-sets used in their work. In any case, it is an encouraging result that these stars may be used as distance indicators even for systems or populations for which the metallicity is not known. However, more systems with DC stars coming from different environments (age, metallicity, star formation histories) are needed in order of confirming a general use of such PL relationship. In this respect, the dSph galaxies contain stars that inhabit a different age/metallicity range than found  anywhere else in the Galaxy and so offer the chance to study DC (and other variables) from unique and otherwise hard to study populations. 

A previous study in Carina \citep{vivas13} has raised several interesting points: {\sl (i)} Carina is very rich in DC stars. It has the highest specific frequency of DC among the other extra-galactic systems and the globular clusters of the Milky Way; {\sl (ii)} There seems to be a fundamental difference between the DC population in dSph galaxies and the galactic field. While high amplitude DC are extremely rare in the field \citep{balona12}, they are at least $100\times$ more frequent in Carina. {\sl (iii)} There are important differences observed among the properties of the DC population of  Carina, Fornax and the LMC which may be a reflection of a metallicity spread, depth along the line of sight, and/or different evolutionary paths of the DC.

In this work we search for DC stars in another of the dSph galaxies around the Milky Way, Sextans. Our main goal is characterize the DC population in this galaxy and increase the number of such stars in extra-galactic systems. We will defer the derivation of a unique PL relationship for a future paper once other galaxies in our program have been surveyed. 

Sextans is a low luminosity system located at 86 kpc from the Sun, similar to the distance to Carina. But, contrary to Carina, it has an older population and  a much lower surface brightness.  It does not have a strong intermediate age population although there is evidence of continuous star formation in the last $\sim 8$ Gyrs \citep{lee09} and a significant spread in metallicity \citep{battaglia11}. Numerous stars in the blue straggler region have been observed in Sextans \citep{mateo91,mateo95,lee09}. These are the stars that are located in the instability strip and may be pulsating as DC. The search for variable stars in Sextans has been limited in the past due in part to its large extension in the sky \citep[its tidal radius is $\sim 83\arcmin - 120\arcmin$,][]{irwin95,roderick16,okamoto17,cicuendez18}. Prior to this study, no DC stars were known in this galaxy, and even the census for brighter variable stars such as RR Lyrae stars was incomplete \citep{mateo95,amigo12,medina18}. Here we take advantage of the large field of view (FoV) of the Dark Energy Camera \citep[DECam,][]{flaugher15} to search for variable stars in a large extension of the Sextans dSph. 

This paper is structured as follow. In \S\ref{sec:obs} we describe the observational strategy to find DC stars in Sextans and discuss the methods used for reducing the data and obtaining photometry. We search for variable stars in \S\ref{sec:var}. Although the focus of this work is the DC stars, other types of pulsating stars such as RR Lyrae stars and Anomalous Cepheids were also found and characterized. The large spatial coverage in our survey allow us to explore the existence of possible extra-tidal variables in Sextans (\S\ref{sec:gaia}). The distance to Sextans from its RR Lyrae stars is derived in \S\ref{sec:distance}. Then, in \S\ref{sec:PL-DC}, we derive a PL relationship for DC stars in Sextans by anchoring to the distance modulus obtained by the RR Lyrae stars. Finally, we discuss and compare the properties of DC stars in Sextans with the ones known in other extra-galactic systems in \S\ref{sec:propertiesDC}. Concluding remarks are given in \S\ref{sec:conclu}.

\section{Observations} \label{sec:obs}

\floattable
\begin{deluxetable}{ccclcc}[htb!]
\small
\tablecolumns{6}
\tablewidth{0pc}
\tablecaption{Observing Log \label{tab:observations}}
\tablehead{
Field  & $\alpha$(J2000.0) & $\delta$ (J2000.0) & Date & $N_g$ & $N_r$ \\
}
\startdata
A & 10:09:12.0 & -02:21:01.4 & 2014 March 8-10 & 34 & 35 \\
B  & 10:17:12.0 & -00:50:35.5 & 2014 March 8-10 & 38 & 38 \\
C  & 10:13:03.0 & -01:31:53.0 & 2017 Apr 4 & 48 & 4 \\
\enddata
\end{deluxetable}

\begin{figure}[htb!]
\plotone{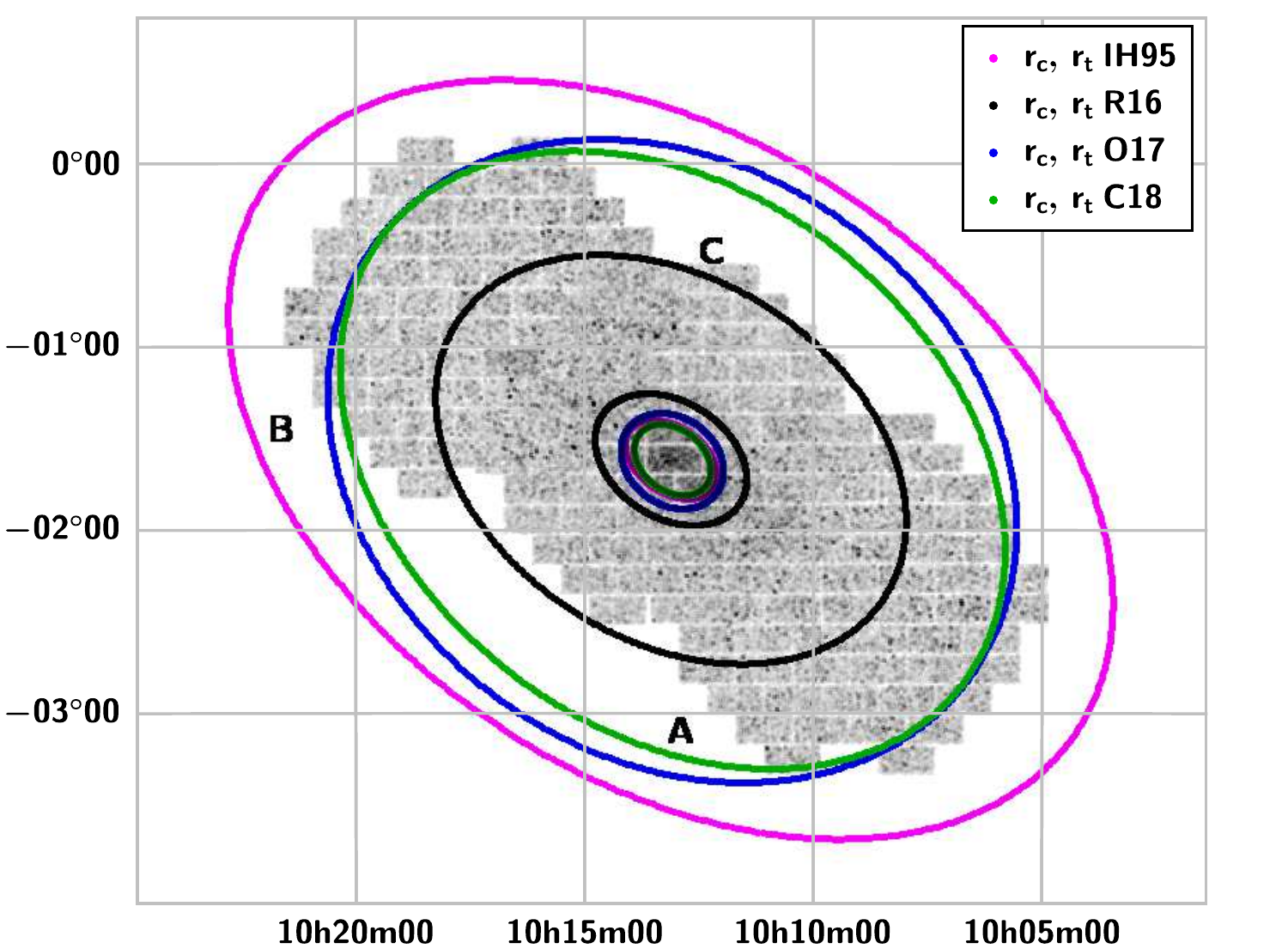}
\caption{Density map in equatorial coordinates of objects detected in the three fields observed with DECam, labeled A, B and C (Table~\ref{tab:observations}). The ellipses with different colors indicate the King's core radius ($r_c$) and tidal radius ($r_t$) determined by \citet[IH95,][magenta]{irwin95}, \citet[R16,][black]{roderick16}, \citet[O17,][blue]{okamoto17}, and \citet[][green]{cicuendez18}.}
\label{fig:footprint}
\end{figure}

DC stars are faint pulsating stars located in the instability strip, $\sim$1.5 to 2.5 magnitudes below the HB. Since the Sextans HB is located at $g\sim 20.5$, we expect the Sextans DC stars to have magnitudes $g$ between 22 and 23. The other important property of DC stars necessary for planning the observational campaign is their pulsational periods. They have very short periods, $\lesssim 0.1$ days, with peak close to 0.06 days, or 1.4 hours \citep{breger00}. In consequence, for distant systems such as Sextans, medium/large  aperture telescopes are required since exposure times should be kept short so they do not cover a significant fraction of the pulsation period.

DECam is an ideal instrument to pursue a survey for DC stars in Sextans. Not only it has a large FoV (3 sq degrees) but also it is installed in a 4m telescope, the Blanco Telescope at Cerro Tololo Inter-American Observatory, Chile. We observed 3 fields with DECam which cover a large extension of the galaxy (Figure~\ref{fig:footprint}). The determination of the size of the Sextans galaxy has been quite controversial. In Figure~\ref{fig:footprint} we show the location of the observed DECam fields together with the King's core radius ($r_c$) and tidal radius ($r_t$) of Sextans determined by different works. Using DECam, \citet{roderick16} suggested a tidal radius of $83\farcm 2 \pm 7\farcm 1$, which is significantly smaller than the traditional value of $160\arcmin \pm 50\arcmin$ by \citet{irwin95}. Very recently, \citet{cicuendez18}, also using DECam, and \citet{okamoto17}, with Suprime-Cam, challenged \citet{roderick16}'s findings by measuring a tidal radius of $\sim 120\arcmin$. If assuming \citet{roderick16}'s tidal radius, our survey covers virtually the full extension of the Sextans.  If the true $r_t$ is closer to the value given by \citet{okamoto17} and \citet{cicuendez18}, our survey covers the galaxy completely along the semi-major axis, but it is incomplete along the semi-minor axis.

Observations were taken during 2 observing runs in 2014 and 2017.  Table~\ref{tab:observations} summarizes the DECam observations. The advantage of the short periods of DC stars is that the light curves can be fully sampled with a few hours of observations.  Our strategy consisted in obtaining continuous observations of our field in two bands ($g$ and $r$) for several consecutive hours each night. In 2014, we covered 2 fields at each side of the center of Sextans, placed along the semi-major axis of the galaxy (Fields A and B, see Figure~\ref{fig:footprint}). In each of the first two nights of the 3-night run we gave priority to a different field in order of obtaining a set of continuous $g,r$ exposures during $\sim 6$ hours, and secure the observation of at least 1 full pulsational period of the DC stars. Nonetheless, a few observations of the other field were inserted through the night. The observing sequence consisted in $300-500$s in $g$ and $r$, respectively. The third night we alternated between the 2 fields continuously. In this observing run we collected $\sim 36$ epochs per band per field. Although the cadence was designed for DC stars, having observations during 3 nights made the data-set suitable for longer period variables such as RR Lyrae stars and Anomalous Cepheids.

For the 2017 run we had only one night available so the observing strategy changed and we prioritized the number of epochs over multi-band observations. The goal for this run was to cover the central part of the galaxy (Field C). We ran sequences of $10 \times 300$s in $g$ and $1 \times 500$s in r, consecutively for 5.5h, the time in which the galaxy was above an airmass of $\sim 1.8$.  
In total we obtained 48 and 4 observations in $g$ and $r$ respectively. We note that CCD S30 was active during this run (contrary to the 2014 observing run).
The seeing was stable and similar for both runs, with a median value of $1\farcs 26$. The moon was closer to Sextans during the 2017 observations. Consequently, the limiting $g$ magnitude for the central field C is $\sim 0.5$ mag brighter than for fields A and B ($0.25$ mags brighter in the $r$ band). Data are available through the NOAO Science archive\footnote{\url http://archive.noao.edu}.

\subsection{Data Reduction and Photometry} \label{sec:photometry}

Basic reduction of the data was done by the DECam Community Pipeline \citep{valdes14} which includes a refinement of the WCS defined in the headers of the images. 
Point spread function (PSF) photometry was extracted from the images of each of the individual detectors in DECam using the DoPHOT software package \citep{schecter93,alonso12}. The instrumental positions of the sources reported by DoPHOT were transformed into ecliptic coordinates using WCSTools and the astrometric information provided in the headers of the reduced images. Extended objects flagged by DoPHOT were eliminated.

For each field, we chose reference catalogs in both filters based on the number of objects detected, seeing and airmass. Photometry from the different epochs was first brought into the instrumental system defined by this reference catalog. Reference catalogs were first cleaned by eliminating all objects whose photometric error ($\sigma_{rm star}$) was larger than the photometric error of most of the stars of similar magnitude. This process was accomplished by calculating a clipped mean and standard deviation ($\sigma_m, rms$) of the DoPHOT individual errors for all objects in bins of 0.25 mags. A curve of maximum expected error ($\sigma_{\rm exp} (mag)$ was defined as a spline function going through $\sigma_m + 5 \times rms$. Objects that had  $\sigma_{rm star} > \sigma_{\rm exp}$ were eliminated from our working catalog. This process was particularly important to eliminate spurious detections within large galaxies, which were abundant in these fields.

In the next step, all catalogs were matched to the reference catalogs using STILTS\footnote{\url http://www.star.bris.ac.uk/$\sim$mbt/stilts/} \citep{taylor06} with a tolerance of $0\farcs 7$. Only objects having a minimum of 12 observations in both $g$ and $r$ were kept. For Field C which was observed very few times in the $r$ band, the minimum number was set to 2 in that band. Each catalog was normalized to the reference by calculating zero point differences in each filter using all stars in common brighter than $g,r = 20$ mag, in each CCD.  Typically there were $\sim 80-100$ stars per CCD available for this calculation and the resulting rms was usually $<0.02$ mag. The zero points calculated this way were applied to all stars in each photometric catalog.

\begin{figure}[htb!]
\centering
\includegraphics[width=7.5cm]{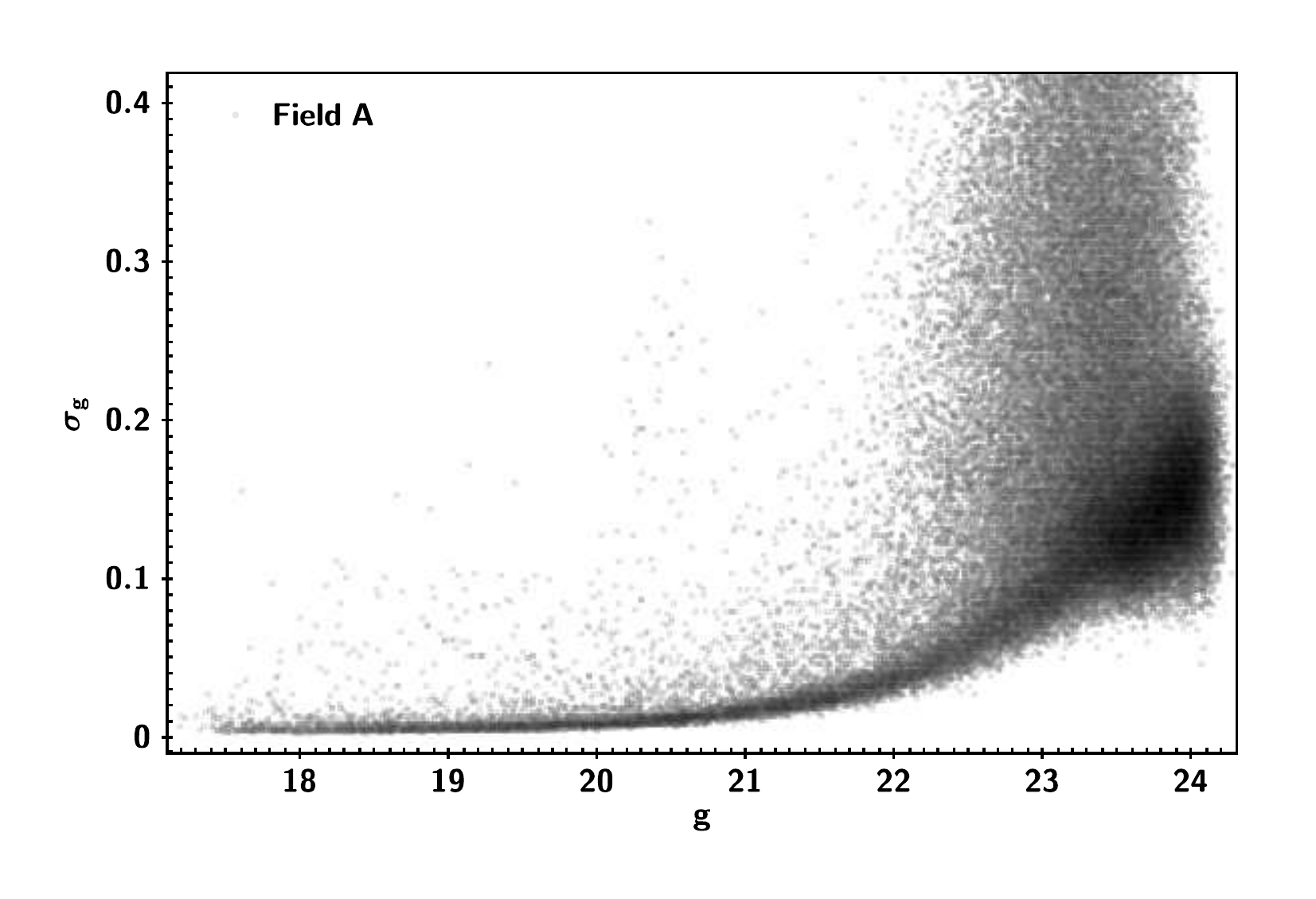}
\includegraphics[width=7.5cm]{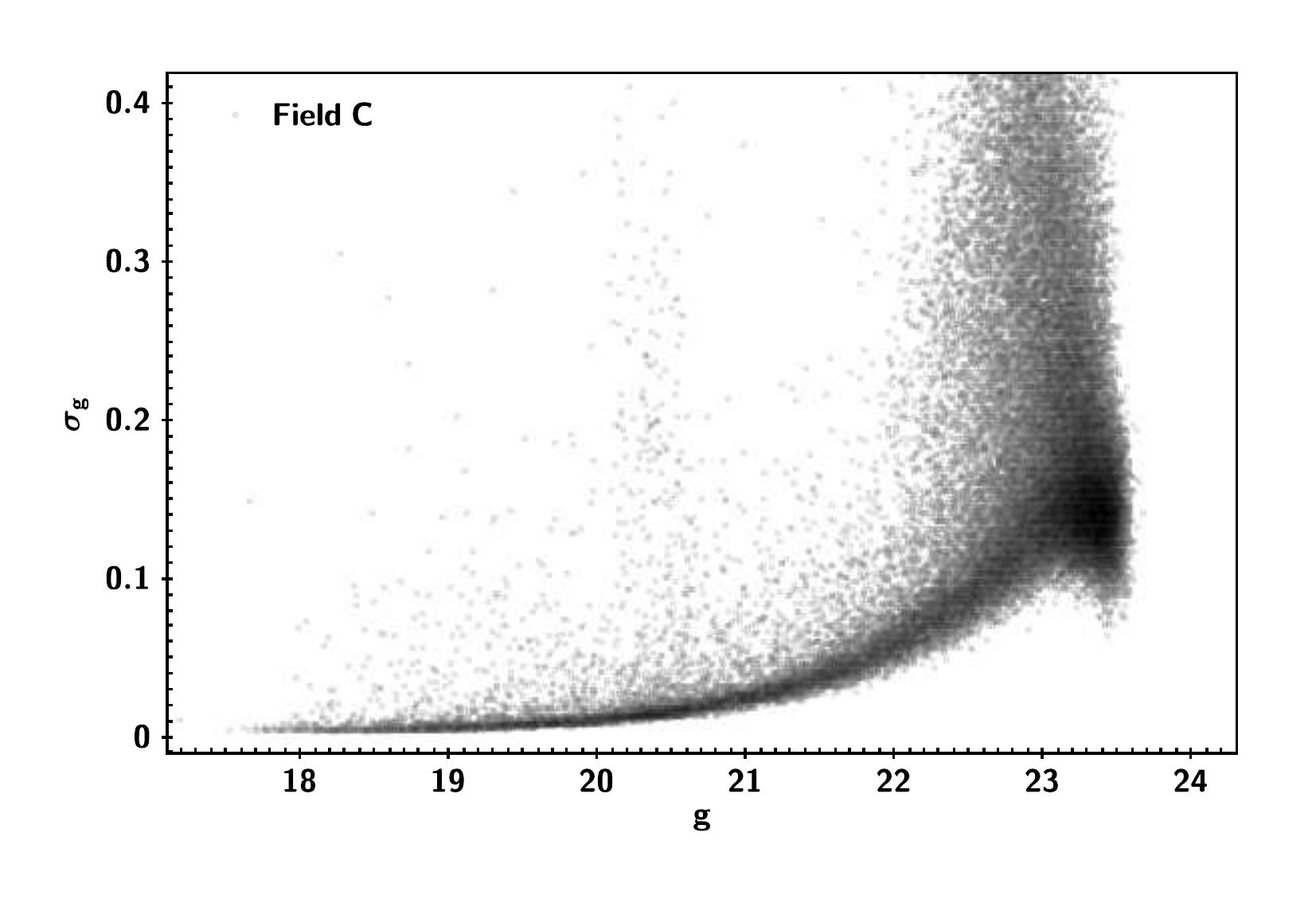}
\caption{Standard deviation of the magnitude distribution in $g$ for all stars in Field A (Top) and Field C (Bottom) as a function of mean magnitude.}
\label{fig:error}
\end{figure}

In Figure~\ref{fig:error} we show the standard deviation of the magnitude distribution for each star as a function of mean 
magnitude for fields A and C. Field B, not shown,  is very similar to A. Errors increase to 0.1 mag at $g=23.1$ in Fields A and B, and at $g=22.6$ in Field C. Variable stars are recognized in a plot like this because the standard deviation of the magnitude distribution is larger than the main locus observed in Figure~\ref{fig:error}, which correspond to stars that keep a constant magnitude within their photometric error. To characterize the loci of non-variable stars we binned the data in 0.25 mag bins and calculated the (sigma-clipped) mean error, $\sigma (m)$, and standard deviation, ${\rm{std}}(m)$, in each bin centered on magnitude m.  We consider stars as variable candidates if they are located significantly above the locus (${\sigma }_{{\rm{star}}} \geq \sigma (m)+3\times {\rm{std}}(m)$) in both the $g$ and $r$ filters.
In Figure~\ref{fig:error}, particularly for Field C, is very obvious the number of variable star candidates at $g\sim 20.5$. These stars correspond to the magnitude of the horizontal branch in Sextans which is populated by numerous RR Lyrae stars.

Absolute calibration was made using Pan-STARRS1 (PS1) DR1\footnote{\url https://panstarrs.stsci.edu} photometry \citep{magnier16}. Each field was calibrated separately. We matched stars with PS1 with a tolerance of $1\arcsec$, selecting only stars with PS1 photometric errors $<0.05$ mag. Linear transformation equations (zero point and color term) between our instrumental magnitudes and the PS1 stars were calculated for each field using stars from all CCDs. We found no need for doing this step in a CCD by CCD basis since the results were similar for all CCDs, but errors are minimized having a larger number of calibrating stars when using the full field. Between 2,000 and 3,500 stars were used for the calibration in each field. The rms of the fits are 0.02 mag in both $g$ and $r$.

\section{Variable Stars in the Instability Strip} \label{sec:var}

Stars flagged as variable candidates, as described in the previous section, were searched for periodicity using a multi-band implementation of the \citet{lafler65} algorithm, as described in \citet{vivas13}. The method is a phase dispersion minimization algorithm in which the correct period is the one that produces a smooth phased light curve. To optimize the search, we imposed a color constrain in order to include only variable stars candidates in the instability strip. The color cut was loose enough, $-0.5 < (g-r) < 0.62$, to include a generous region around the instability strip but avoiding to search through the large number of very red sources present in the color-magnitude diagram (CMD). To search for DC stars we used trial periods ranging from 0.03 to 0.15 days, while for RR Lyrae stars and Anomalous Cepheids, the range for searching was 0.15 to 0.9 days. The limited time baseline of the data does not allow to search for periodicity beyond 1 day. All stars passing the cut $\Lambda>2.0$ as defined by  \citet{lafler65} were visually inspected. The parameter $\Lambda$ quantifies the significance of the period selected as the best. In order to account for possible aliasing and spurious periods, the three best periods within the search range were inspected. Periodic variable stars were finally selected during this inspection and classification was made based on the properties of the light curve (period and amplitudes) and their position in the CMD. Four stars had periods and lightcurves in agreement with them being either RR Lyrae stars or DC but they were not in the corresponding place in the CMD for their type. Most likely those stars are not Sextans members but just field stars in the foreground of the galaxy (but see discussion in \S\ref{sec:gaia}).

Because there is overlap between fields, some variables were recognized independently in each field which gave us confidence in our selection method. In those cases, we chose the detection in Fields A and B over the ones in C since the latter does not have time series in $r$. 

\begin{figure}[htb!]
\epsscale{1.2}
\plotone{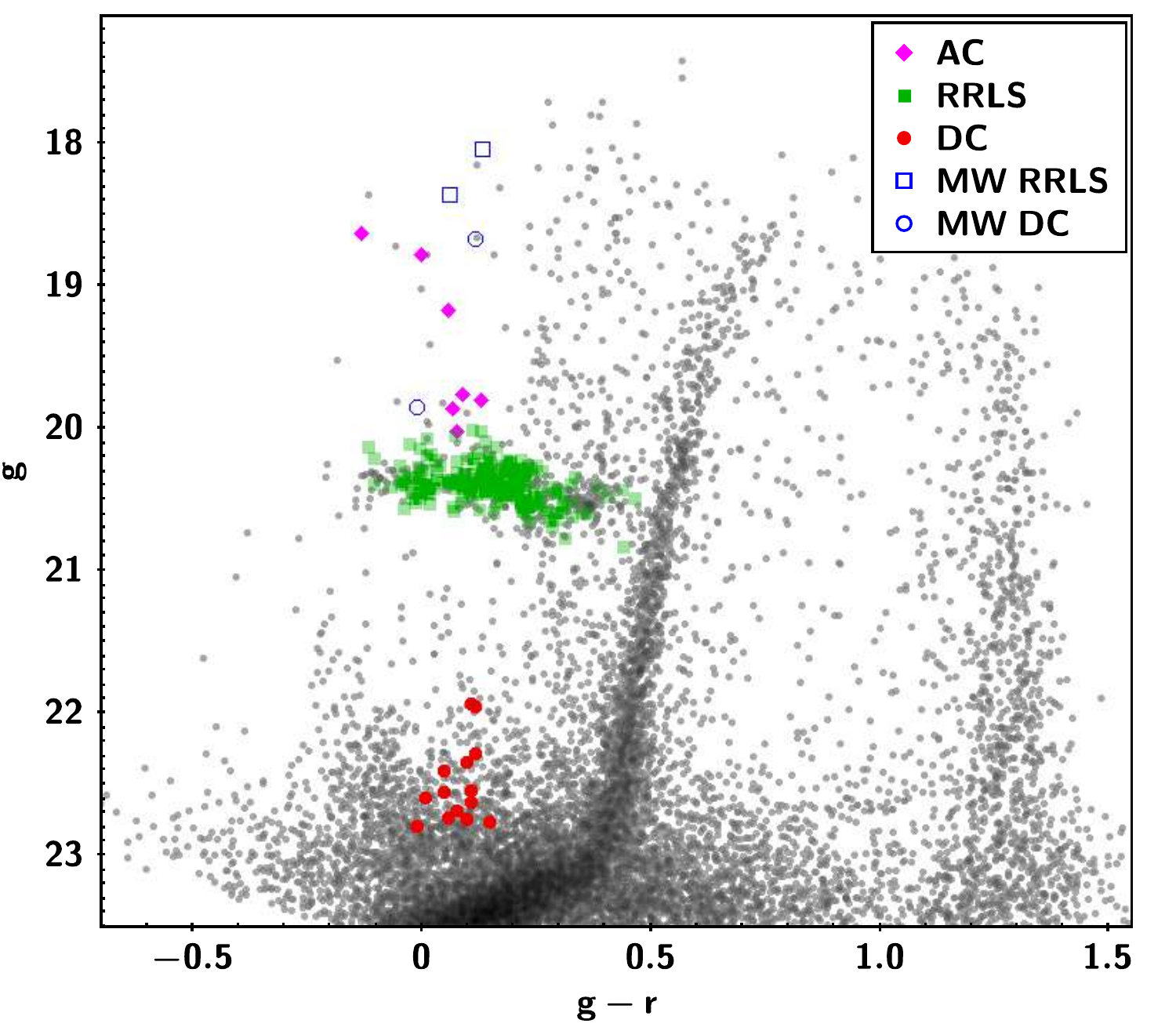}
\caption{Color magnitude diagram of Sextans. To better see the features of the galaxy we only included stars within the core radius ($26\farcm 8$) of the galaxy. Pulsating variable stars detected in the full footprint of the survey are overplotted in the CMD with symbols as indicated in the legend.}
\label{fig:CMDvar}
\end{figure}

\begin{figure*}[htb!]
\epsscale{0.9}
\plotone{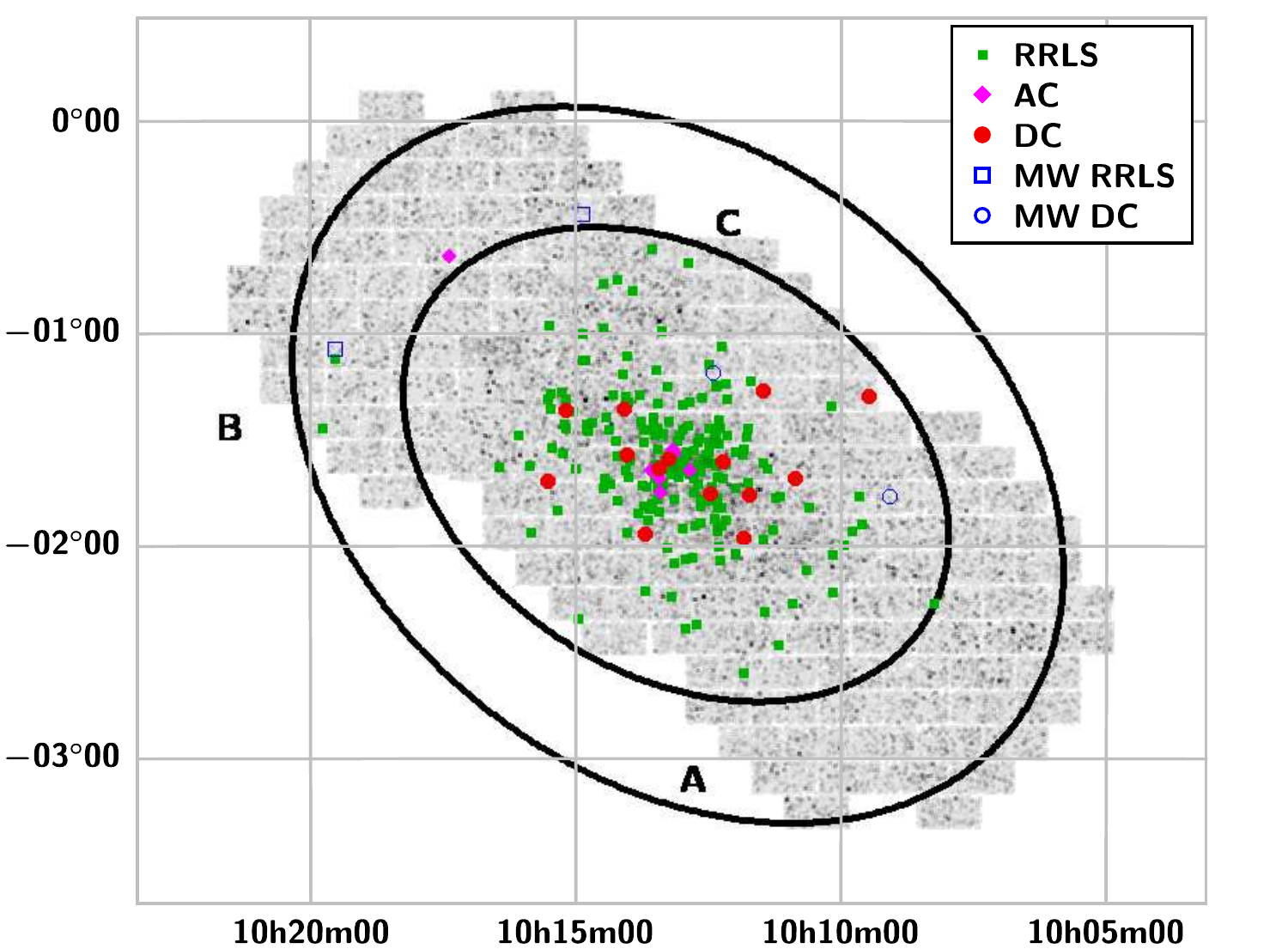}
\caption{Map in equatorial coordinates of the main groups of variable stars found in this work. The inner and outer ellipses indicate a tidal radius of $83\farcm 2$ \citep{roderick16} and $r_t=120\arcmin$ \citep{okamoto17,cicuendez18}, respectively.}
\label{fig:sky-var}
\end{figure*}

Figure~\ref{fig:CMDvar} shows the CMD of Sextans with the variable stars identified. To better see the features of the Sextans galaxy we limited the stars in the CMD to those inside an ellipse with a semi-major axis equal to the core radius of Sextans \citep[we assumed the parameters derived by][including an ellipticity of 0.29 and Position Angle of 56$_{.}^{\circ}$7]{roderick16}. The narrow red giant branch and sub-giant branch of Sextans are clearly defined in this diagram. The HB contains numerous RR Lyrae stars. Figure~\ref{fig:sky-var} shows the spatial distribution of the different types of variables. In the next subsections we discuss the main types of variables found in Sextans, from fainter to brighter.

\subsection{Dwarf Cepheids} \label{sec:DC}

\begin{deluxetable*}{cccccccccccc}[htb!]
\tablecolumns{12}
\tabletypesize{\scriptsize}
\tablewidth{0pc}
\tablecaption{Dwarf Cepheid Stars \label{tab:DC}}
\tablehead{
ID  & $\alpha$(J2000.0) & $\delta$ (J2000.0) & $N_g$ & $N_r$ & $\langle g \rangle$ & $\langle r \rangle$ & Period &
$\Delta_g$  & $\Delta_r$ & $D_{\rm Sex}$  & Comment \\
      & (deg) & (deg) & & & (mag) & (mag) & (d) & (mag) & (mag) & ($\arcmin$) & \\
}
\startdata
 DC1 & 152.37167 & -1.29472 & 47 & 4 & 22.59 & 22.58 & 0.0591 & 0.99 & ... & 56.5 &   Sex \\ 
 DC2 & 152.71577 & -1.68032 & 47 & 4 & 22.34 & 22.24 & 0.0631 & 0.58 & ... & 33.1 &   Sex \\ 
 DC3 & 152.86590 & -1.27122 & 46 & 4 & 21.93 & 21.82 & 0.0853 & 0.70 & ... & 30.9 &   Sex \\ 
 DC4 & 152.92910 & -1.75998 & 34 & 35 & 22.54 & 22.43 & 0.0576 & 0.89 & 0.54 & 22.2 &   Sex \\ 
 DC5 & 152.96042 & -1.96451 & 32 & 35 & 22.74 & 22.68 & 0.0521 & 0.91 & 0.60 & 28.4 &   Sex \\ 
 DC6 & 153.05480 & -1.60169 & 48 & 4 & 22.76 & 22.61 & 0.0583 & 1.14 & ... & 12.5 &   Sex \\ 
 DC7 & 153.11685 & -1.75687 & 48 & 4 & 21.96 & 21.84 & 0.0783 & 0.68 & ... & 12.8 &   Sex \\ 
 DC8 & 153.31278 & -1.59130 & 43 & 4 & 22.62 & 22.51 & 0.0604 & 0.95 & ... &  3.1 &   Sex \\ 
 DC9 & 153.35546 & -1.63900 & 46 & 4 & 22.28 & 22.16 & 0.0914 & 0.79 & ... &  6.0 &   Sex \\ 
DC10 & 153.42480 & -1.94532 & 45 & 4 & 22.79 & 22.80 & 0.0509 & 0.94 & ... & 22.9 &   Sex \\ 
DC11 & 153.50747 & -1.56894 & 45 & 4 & 22.56 & 22.51 & 0.0662 & 0.83 & ... & 14.8 &   Sex \\ 
DC12 & 153.52053 & -1.35718 & 41 & 4 & 22.75 & 22.65 & 0.0543 & 0.78 & ... & 21.3 &   Sex \\ 
DC13 & 153.79662 & -1.36580 & 36 & 38 & 22.68 & 22.60 & 0.0588 & 0.91 & 0.64 & 35.0 &   Sex \\ 
DC14 & 153.87762 & -1.69308 & 36 & 38 & 22.41 & 22.36 & 0.0679 & 1.10 & 0.84 & 37.3 &   Sex \\ 
DC15 & 152.26788 & -1.77082 & 34 & 35 & 18.67 & 18.55 & 0.0769 & 0.51 & 0.37 & 60.5 & Field \\ 
DC16 & 153.10447 & -1.18371 & 48 & 4 & 19.85 & 19.86 & 0.0551 & 0.67 & ... & 26.7 & Field \\ 
\enddata
\end{deluxetable*}

\begin{figure*}[htb!]
\epsscale{0.8}
\plotone{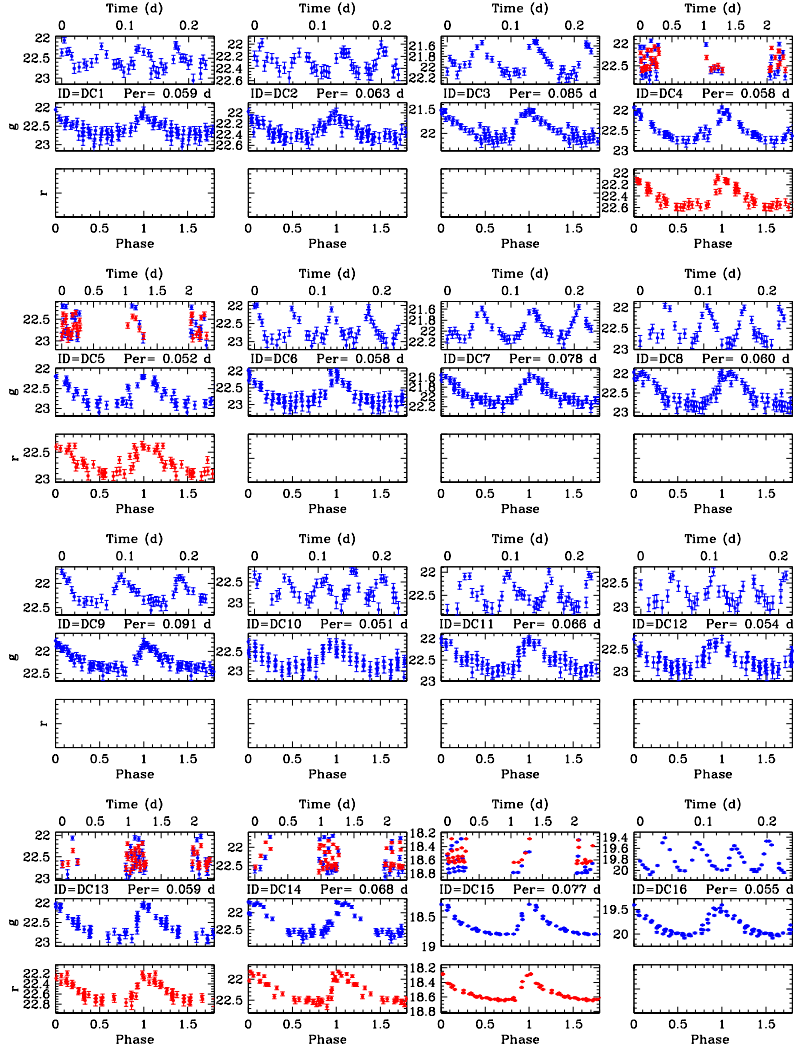}
\caption{Light curves of the 16 DC stars found in Sextans. Magnitudes in $g$ and $r$ are shown with color blue and red, respectively. The top panel for each star is the time series in days from the time of the first observation. The two lower panels show the phased light curve. Stars DC-15 and DC-16 are likely field stars.}
\label{fig:lcDC}
\end{figure*}

DC were the main goal of this work. A total of 16 DCs were detected, although 2 of them are significantly brighter than the rest and above the HB. It is very likely that these two DC stars are actually field stars (see Figure~\ref{fig:CMDvar}), although it is still puzzling that they are both inside the tidal radius of Sextans (Figure~\ref{fig:sky-var}).
Figure~\ref{fig:lcDC} shows the lightcurves of the DC stars. For each star we show the time series (top panels), and the phased light curves in both $g$ and $r$ (lower panels). The stars with only $g$ data correspond to the ones detected in Field C. The time span of the observations in Field C was only 5.5 hours but the top panels clearly show that it was possible to cover 3-4 pulsational cycles and hence, we could easily determine the periods. The position, light curve properties (mean magnitudes, period and amplitudes) and distance to the center of Sextans ($D_{Sex}$) of these DC stars are recorded in Table~\ref{tab:DC}. The last column in the Table indicates if they belong to the Sextans galaxy or if they are field stars. The reported mean magnitudes in Table~\ref{tab:DC} are not plain averages but phase-weighted intensity-averaged magnitude calculated following \citet{saha90}. This way to calculate the mean magnitudes avoid the biases toward minimum magnitudes (because as RR Lyrae stars, some DC stars also spend most of their pulsation cycle time at minimum light) and those biases that may appear in unevenly sampled lightcurves. The reported $r$ magnitudes for stars in field C are straight average of the few (4) available measurements. Consequently, colors for these stars may not be the true mean color. 

\begin{figure}[htb!]
\plotone{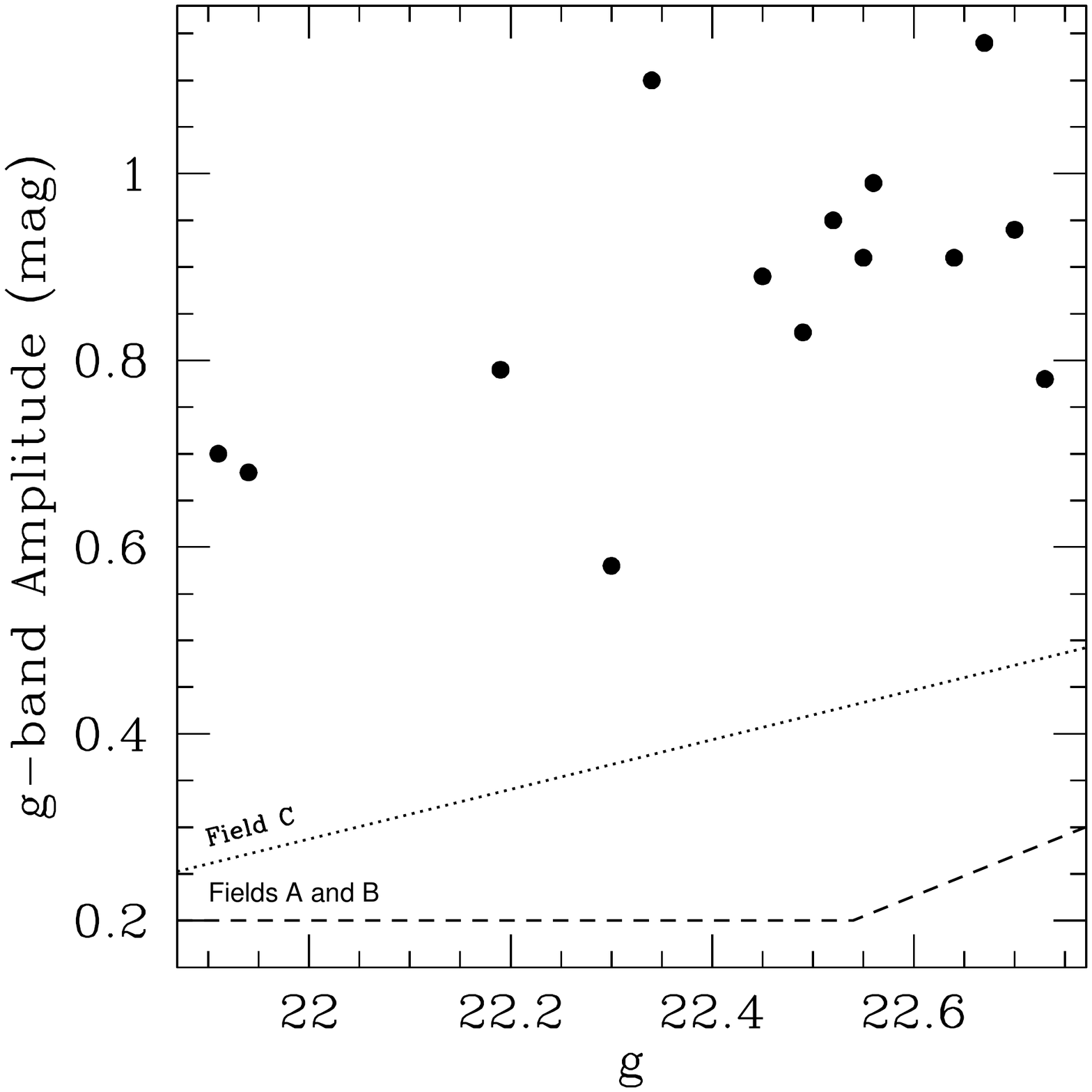}
\caption{Amplitude of the DC stars as a function of mean magnitude. The dashed and dotted lines indicate the expected minimum amplitude that can be measured in Fields A (similar to Field B) and Field C, respectively.}
\label{fig:amp}
\end{figure}

The DC stars in Sextans occupy a narrow color range in the CMD, $\Delta (g-r) = 0.2$ mag, but display a large spread in magnitude, over $\sim 0.8$ mag in $g$, between 21.9 and 22.7 mag. Periods range from 0.05 to 0.09 days, with a mean of 0.0646 d. Most of the lightcurves shown in Figure~\ref{fig:lcDC} look similar to the asymmetrical type ab RR Lyrae stars, although a few ones seem more sinusoidal. However, it is well known that pulsation mode cannot be recognized in DC stars based neither on the lightcurve shape nor in a Period-Amplitude diagram \citep[eg.][]{vivas13}. Thus, the asymmetrical lightcurves in this case do not necessarily mean that those DC stars  are pulsating in the fundamental mode. The amplitudes of variation displayed by the DC stars is large, ranging from $\Delta_g$ = 0.58 to 1.14 mags, with a mean of 0.87 mags. It is possible however that given our photometric errors ($\sim 0.1$ mag at $g=22.6$ in the central Field C) we are missing lower amplitude variable stars. In Figure~\ref{fig:amp} we show the $g$-band amplitude of the DC stars as a function of mean $g$ magnitude.
Based in our past experience in Carina \citep{vivas13}, it should be possible to detect variables with amplitudes of $>0.2$ mags if the photometric errors are $\lesssim 0.05$ mag. When errors increase to $\sim 0.09$ mags, the minimum amplitude that can be detected is $\sim 0.4$ mags. Accordingly, the dotted and dashed lines in Figure~\ref{fig:amp} show the minimum amplitude that can be detected in our survey. Although, we indeed may be missing low-amplitude variables, particularly in the central Field C, it is nonetheless surprising that the lowest amplitude of our DC stars is $\sim 0.6$ mag, well above our detection limit. Galaxies like Carina or the LMC have plenty of stars with amplitudes in the range 0.2-0.6 mags \citep{garg10,vivas13}.

A comparison of the properties of DC stars in Sextans with other extra-galactic systems will be discussed later in this paper.

\subsection{RR Lyrae stars} \label{sec:RRLS}

\begin{splitdeluxetable*}{cccrcccccccrBcccccrrr}
\tablecolumns{20}
\tabletypesize{\scriptsize}
\tablewidth{0pc}
\tablecaption{RR Lyrae Stars \label{tab:RR}}
\tablehead{
ID  & $\alpha$(J2000.0) & $\delta$ (J2000.0) & Type & Period & MJD$_0$ & $N_g$ &  $\Delta_g$ & $\langle g \rangle$ & $g_{\rm max}$ & $\sigma_{\rm fit} (g) $ & Template ($g$)  & $N_r$ &  $\Delta_r$ & $\langle r \rangle$ & $r_{\rm max}$ & $\sigma_{\rm fit} (r)$ & Template ($r$)  & Other Names  & Comment \\
 & (deg) & (deg) & &  (d) & (d) & & (mag) & (mag) & (mag) & (mag) & & (mag) & (mag) & (mag) & (mag) & & & \\
}
\startdata
    RR1 & 152.06381 & -2.27372 &  ab & 0.6948 & 56726.30554 & 32 & 0.94 & 20.29 & 19.74 & 0.02 & 111g.dat & 35 & 0.67 & 20.12 & 19.75 & 0.01 & 109r.dat &    HiTS100815-021625 & Sextans \\ 
  RR2 & 152.40204 & -1.89860 &   c & 0.3457 & 56724.17884 & 34 & 0.61 & 20.36 & 20.08 & 0.07 &   0g.dat & 34 & 0.46 & 20.31 & 20.09 & 0.06 &   1r.dat &    HiTS100936-015356 & Sextans \\ 
  RR3 & 152.41282 & -1.76577 &  ab & 0.7487 & 56726.09288 & 34 & 0.56 & 20.30 & 19.98 & 0.01 & 101g.dat & 35 & 0.42 & 20.08 & 19.86 & 0.01 & 104r.dat &                           & Sextans \\ 
  RR4 & 152.44702 & -1.93215 &   c & 0.2987 & 56725.25898 & 34 & 0.23 & 20.30 & 20.19 & 0.01 &   0g.dat & 34 & 0.16 & 20.28 & 20.21 & 0.01 &   0r.dat &                           & Sextans \\ 
  RR5 & 152.48994 & -1.99834 &  ab & 0.6553 & 56725.10121 & 33 & 0.86 & 20.38 & 19.90 & 0.04 & 101g.dat & 33 & 0.74 & 20.19 & 19.79 & 0.03 & 109r.dat &    HiTS100958-015954 & Sextans \\ 
\enddata
\tablecomments{Table~\ref{tab:RR} is published in its entirety in the machine-readable format. A portion is shown here for guidance regarding its form and content.}
\end{splitdeluxetable*} 

\begin{figure}[htb!]
\epsscale{1.2}
\plotone{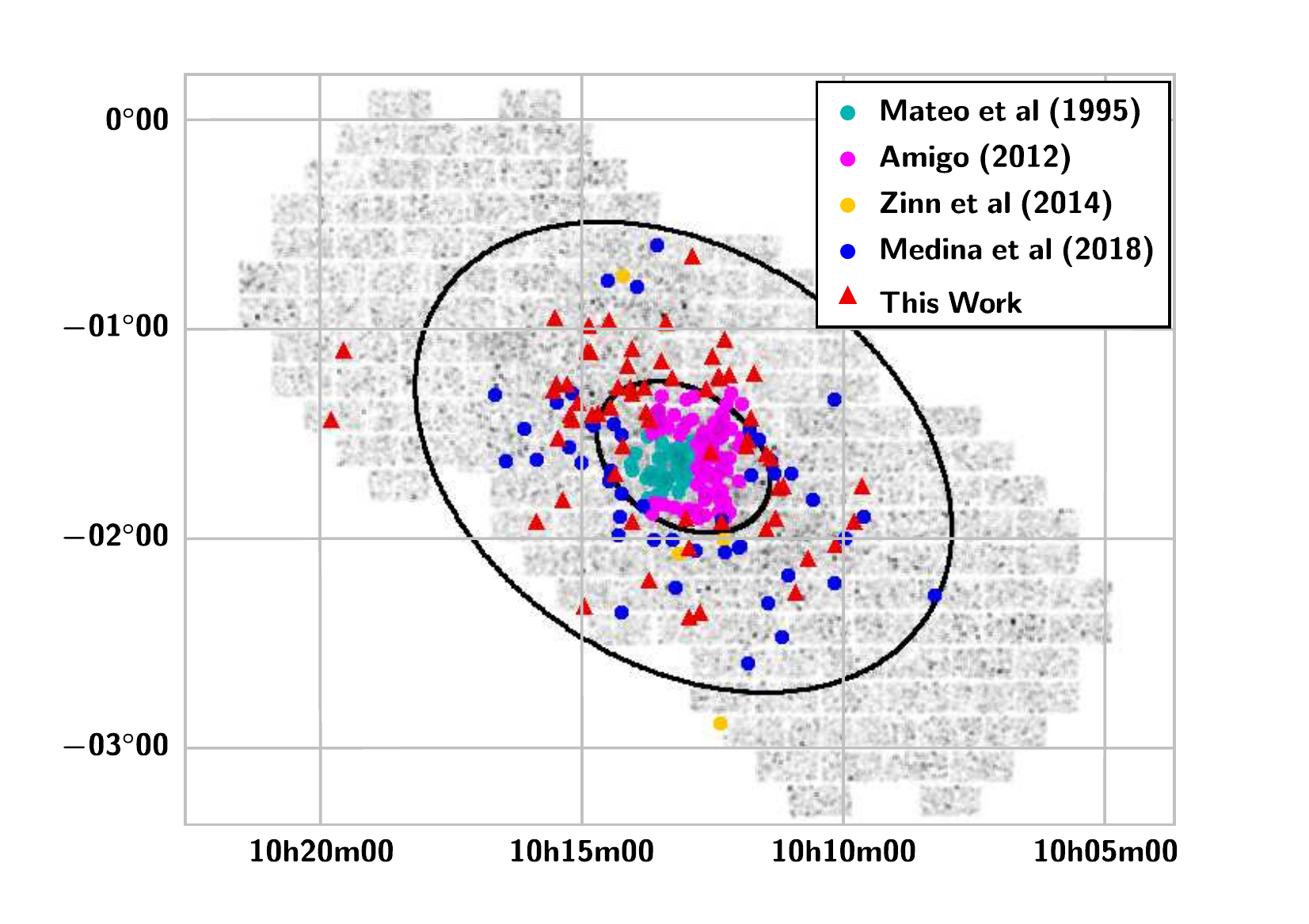}
\caption{Distribution in the sky of RR Lyrae stars in the Sextans dSph. The color code indicates the first work that identified each RR Lyrae star. The two ellipses correspond to the core radius ($r_c=26.8'$) and tidal radius ($r_t=83.2'$) as determined by \citet{roderick16}. Although most of the RR Lyrae stars are concentrated toward the center of the galaxy, there are several in the outskirts. In particular, there are 3 RR Lyrae stars outside the tidal radius (as determined by \citet{roderick16}), one from La Silla-QUEST survey \citep{zinn14} and two from this work. }
\label{fig:skyRRLS}
\end{figure}

\begin{figure*}[htb!]
\epsscale{0.8}
\plotone{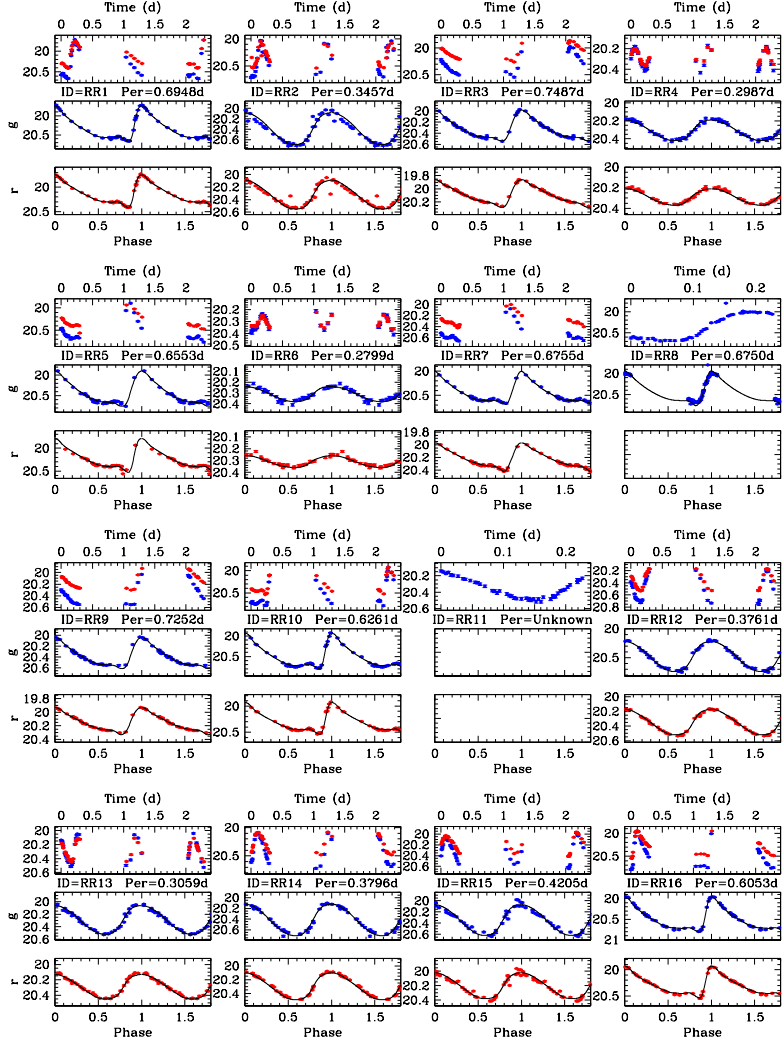}
\caption{Light curves of the RR Lyrae stars detected in this work. The top panel for each star is the time series in days from the time of the first observation. This encompass 3 consecutive nights for some stars (those in fields A and B), and only 5.5 hours (or $\sim 0.23$d) for stars in Field C.  For the later group, only $g$ time series are available, and phase light curves (two lower panels for each star) are shown only for stars with previously known periods (see text). The complete figure set (13 images) is available in the online journal.}
\label{fig:lcRR}
\end{figure*}

RR Lyrae stars are by far the most numerous group of variable stars in Sextans. We detected 199 RR Lyrae variables, 65 of which are new discoveries and 2 are likely foreground stars. RR Lyrae stars have been detected in Sextans before by \citet{mateo95} and \citet{amigo12}, although their studies only cover the central part of the galaxy (Figure~\ref{fig:skyRRLS}). \citet{lee03} had also previously selected RR Lyrae star candidates in an area similar to \citet{amigo12} based on variability between two different epochs, but clearly, characterization of the light curves was not possible. More recently, two large scale RR Lyrae star surveys covered, partially, the Sextans galaxy. The La Silla-QUEST survey \citep[LSQ,][]{zinn14} covers the full extent of Sextans but the faint limit of that survey is just at the magnitude of the HB in Sextans. Consequently, only a handful of stars were detected. Interestingly, there is 1 star from that survey just outside the \citet{roderick16}'s tidal radius of Sextans (although it is inside if considering the tidal radius by either \citet{irwin95} or \citet{cicuendez18}). On the other hand, the High cadence Transient Survey \citep[HiTS,][]{forster16,medina18} studied 3 DECam fields near (but not centered) on Sextans and measured 66 RR Lyrae stars, many of them being new discoveries since their fields were located in the outskirts of Sextans. Joining all these works together, Sextans has 227 RR Lyrae stars (Figure~\ref{fig:skyRRLS}). We missed several of the known RR Lyrae stars due mostly to the gaps between the CCDs since our observations were not dithered. Coordinates, light curve properties and identification in other surveys of all RR Lyrae stars detected in this work is provided in Table~\ref{tab:RR} and their light curves are shown in Figure~\ref{fig:lcRR}. 

For the stars in field A and B, which have good sampling over the full lightcurve in both the $g$ and $r$ filters, we fitted templates to the data in order to better characterize the lightcurve. We used the library of templates set up by \citet{sesar10} based on SDSS Stripe 82 RR Lyrae stars. We first fitted the template in the band containing the larger number of epochs. The fitting was done by $\chi^2$ minimization, covering a range of periods, initial phase, amplitude and magnitude at maximum light around the observed values and the period given by the \citet{lafler65}'s method. We used the resulting period and initial phase as constants to fit the template in the other band. An example of a star for which we applied this procedure is RR1 (see Figure~\ref{fig:lcRR}).

On the other hand, the time baseline of the observations in Field C (the central field) was not adequate for RR Lyrae stars since only covered a fraction of a single pulsation cycle. The search for RR Lyrae stars was done in a different way in this field. In 5.5 h of continuous observations we covered about $\sim 35\%$ of the pulsation cycle of type ab stars and $\sim 75\%$ of the cycle for \rrc. These are long enough ranges for recognizing RR Lyrae stars in a simple look at the Julian date versus magnitude plot. Thus, we looked for continuous variation in the time series of the variable star candidates in the selected color range. No attempt of period finding was performed. However, since the central field of Sextans has been explored before for RR Lyrae stars, we used the known periods of stars in common with other studies and fitted lightcurve templates to the time series. During the fitting process we allowed generous variations on amplitude (around the literature value), initial phase and magnitude at maximum light, but the period was kept constant. Star RR8 (Figure~\ref{fig:lcRR}) is an example of one of the stars in which this procedure was used. 

The mean magnitude reported in Table~\ref{tab:RR} was obtained by integrating the template in intensity units and transforming back to magnitude the result. In average, the rms of the template fits is $\sim 0.03$ mag. The mean $r$ for stars in Field C are straight averages of the available observations.

There were however, several RR Lyrae star candidates (29 stars) in Field C which were not known before and without a knowledge of the period we cannot characterize the light curve.  Their location in the CMD and the smooth variations observed in the 5.5h-period, secure their classification as RR Lyrae stars. We report them too in Table~\ref{tab:RR} and Figure~\ref{fig:lcRR} (see for example star RR 11). Mean magnitudes for the stars in this group are simple averages of the available observations in each band.

Among the Sextans RR Lyrae stars, we were able to classify 125 as \rrab, 41 as \rrc, and 2 as RR{\sl d}. As mentioned before, 29 stars were not classified due to insufficient time coverage. The ab and c groups have mean periods of 0.616d and 0.369d, respectively. The mean period of the {\rrab} stars in Sextans classifies this galaxy in the class Oosterhoff-intermediate, in agreement with previous works \citep{mateo95}. A full discussion on the properties of RR Lyrae stars in Sextans will be deferred to a future paper (Mart{\'\i}nez-V\'azquez et al, in preparation).

Two RR Lyrae stars (RR33 and RR47) have been classified as double mode pulsators, or RR{\sl d}. RR33 was previously known \citep[C9 in][]{amigo12} and our classification is in agreement with the one given by \citet{amigo12}.  No template fitting was possible in those cases. Table~\ref{tab:RR} reports the first overtone periods. Fundamental periods are likely 0.580d and 0.554d for RR33 and RR47 respectively.  Seven additional stars
(RR32, RR39, RR49, RR112, RR128, RR133 and RR144) have been also reported by \citet{amigo12} to be double mode pulsators. However, our limited time coverage data on these stars, all of them in Field C, show them as c-type pulsators, which is the classification we have given in Table~\ref{tab:RR}.

Other two RR Lyrae stars (RR198 and RR199) have mean magnitudes significantly brighter than the horizontal branch of Sextans (see Figure~\ref{fig:CMDvar}). They are also located well outside the \citet{roderick16}'s tidal radius of the galaxy (Figure~\ref{fig:skyRRLS}).  Thus, these 2 stars are likely field stars along the same line of sight of Sextans, although we discuss the possibility that they are actually Anomalous Cepheids of Sextans in \S\ref{sec:gaia}.

\subsection{Anomalous Cepheids} \label{sec:AC}

\begin{figure*}[htb!]
\begin{center}
\includegraphics[scale=0.8,trim=0cm 13cm 0cm 0cm,clip,width=1.0\textwidth]{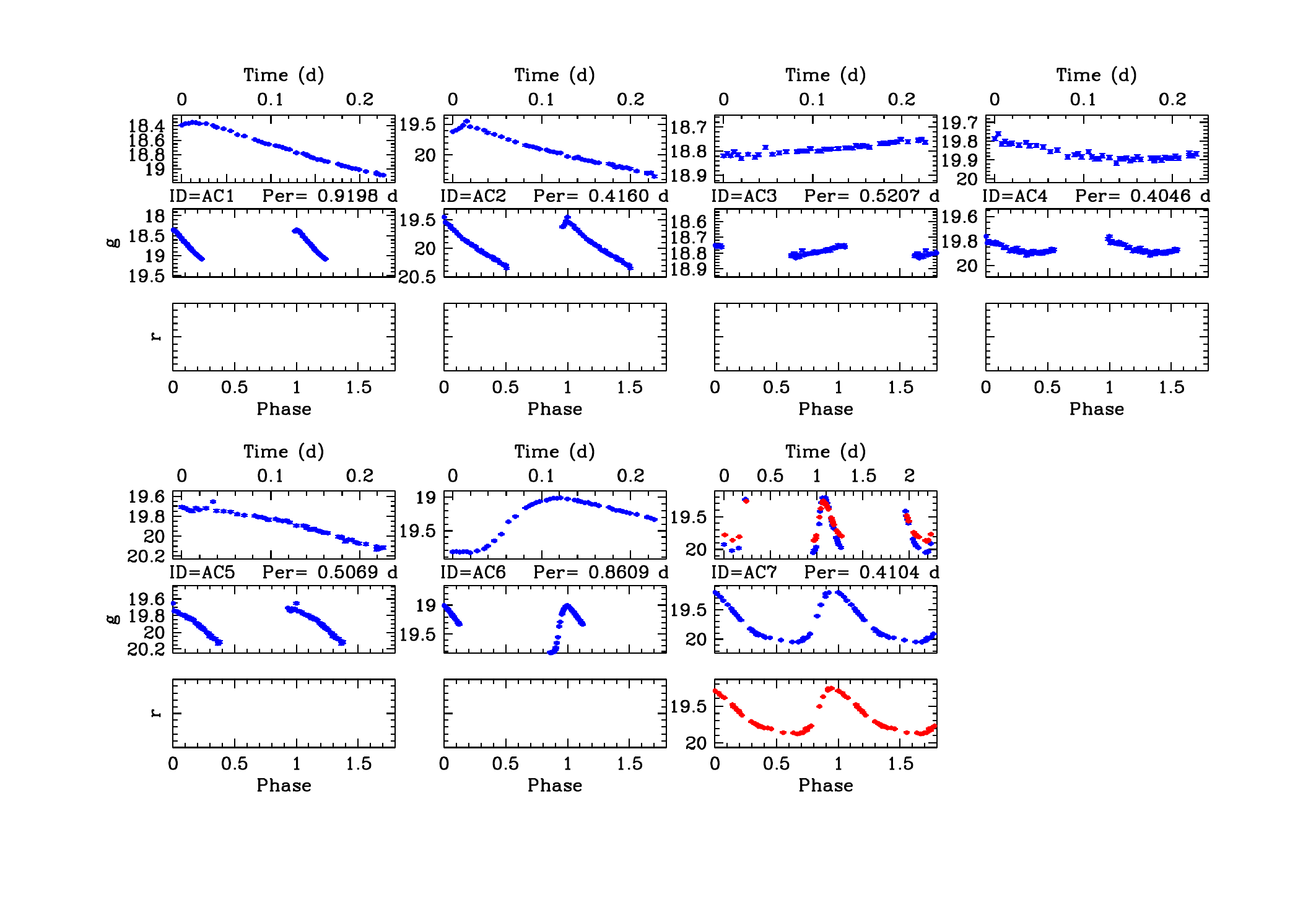}
\caption{Light curves of the 7 Anomalous Cepheid stars detected in this work. The top panel for each star is the time series in days from the time of the first observation. The two lower panels show the phased light curves in $g$ and $r$. Periods for AC1 to AC6 were taken from \citet{mateo95} and \citep{amigo12}. Only for AC7 (a new discovery) we were able to determine the period from our own data. }
\label{fig:lcAC}
\end{center}
\end{figure*}

\begin{deluxetable*}{ccccccccccr}[htb!]
\tablecolumns{11}
\tabletypesize{\scriptsize}
\tablewidth{0pc}
\tablecaption{Anomalous Cepheids \label{tab:AC}}
\tablehead{
ID  & $\alpha$(J2000.0) & $\delta$ (J2000.0) & $N_g$ & $N_r$ & $\langle g \rangle$ & $\langle r \rangle$ & Period &
$\Delta_g$  & $\Delta_r$ & Other Name \\
      & (deg) & (deg) & & & (mag) & (mag) & (d) & (mag) & (mag) & \\
}
\startdata
 AC1  & 153.21128 &   -1.64243 & 46 &  4 &  18.78 &  18.78 & 0.9198 & 0.73 &  ... & V6 \\
 AC2  & 153.28425 &   -1.56633 & 46 &  4 &  20.02 &  19.94 & 0.4160 & 0.90 &  ... &  V9 \\
 AC3  & 153.29410 &   -1.54672 & 46 &  4 &  18.64 &  18.77 & 0.5207 & 0.08 &  ... & C82 \\
 AC4  & 153.35253 &   -1.74856 & 46 &  4 &  19.76 &  19.67 & 0.4046 & 0.16 &  ... & C89 \\
 AC5  & 153.35785 &   -1.68251 & 46 &  4 &  19.86 &  19.79 & 0.5069 & 0.48 &  ... & V34 \\
 AC6  & 153.39361 &   -1.64499 & 46 &  4 &  19.17 &  19.11 & 0.8609 & 0.83  & ...  & V5 \\
 AC7  & 154.34301 &   -0.63833 & 38 & 38 &  19.81 &  19.68 & 0.4104 & 0.85 &  0.62 &  \\
\enddata
\end{deluxetable*} 

Anomalous Cepheids are a common type of variable stars found in dSph galaxies \citep{clementini14}. Their existence in these galaxies is usually interpreted as due to the presence of an intermediate age population since these are metal-poor, 1.3-2.0 $M_\odot$ stars in the core helium burning phase \citep{fiorentino12}. However, an additional mechanism to bring stars to this part of the instability strip is through mass transfer in an old binary system \citep{gautschy17}.  \citet{mateo95} identified 6 Anomalous Cepheids in Sextans, and \citet{amigo12} increased that number to 8. Here, we have recovered 6 of them (the other two were in gaps between the DECam CCDs), and we found an additional star, AC7 in Table~\ref{tab:AC}. Similar to what we did for the DC stars (\S~\ref{sec:DC}), the mean magnitudes in Table~\ref{tab:AC} are phase-weighted intensity-averaged values, except for the $r$ band in field C, which are just straight averages. Because the light curves of these stars are not compeltely sampled, the reported mean magnitudes may not be accurate. The 6 known Anomalous Cepheids were all located in field C and consequently our data does not constraint well the light curve properties. The periods listed in Table~\ref{tab:AC} for these stars comes from \citet{amigo12}. The amplitudes in the table come from our data but they are not well constrained since we do not observe a full cycle for these stars. This is particularly true for AC3 and AC4 for which we cover only part of the lightcurve at minimum light. Our new discovery, AC7, was located in Field B and the period and amplitudes in both $g$ and $r$ are well constrained by our data, and its properties and location in the CMD agree well with the classification as an anomalous Cepheid. 

Surprisingly however, AC7 is not located toward the center of the galaxy as the rest of the Anomalous Cepheid stars in Sextans. Instead, it is located just outside the tidal radius of the galaxy \citep[as measured by][]{roderick16}. We considered the possibility that this star is actually a foreground RR Lyrae star within our footprint. The amplitude and shape of the light curve seem to indicate this may be a {\rrab } star. The period ($0.4104$ d) is somewhat short for a {\rrab } star, but not completely impossible \citep{samus17}. AC7 has a mean $g$ magnitude of 19.72, which will locate this star at $\sim 53$ kpc from the Sun (after correcting by interstellar extinction) if it were an RR Lyrae star. Stars at such large distances in the Galactic halo are rare. We integrated the number density radial profile of RR Lyrae stars derived by \citet{medina18} using a large sample of distant stars, and found that we should expect 0.25 RR Lyrae stars in the range of distances 50-60 kpc in the 7 sq degree area covered by our survey. Thus, it would not be completely impossible this is indeed a foreground RR Lyrae star. However, short period ab-type RR Lyrae stars are usually associated with high metallicity  \citep[for example,][]{skarka14,maintz05}, which would be unusual given its large distance from the Sun and location above the galactic plane ($b\sim 40\degr$).
We discuss more on the spatial distribution of the variables in Sextans  and the possibility of extra-tidal variable stars in the next section.

\section{Spatial Distribution of the variable stars in Sextans} \label{sec:gaia}

\begin{figure*}[htb!]
\centering
\includegraphics[width=0.45\textwidth]{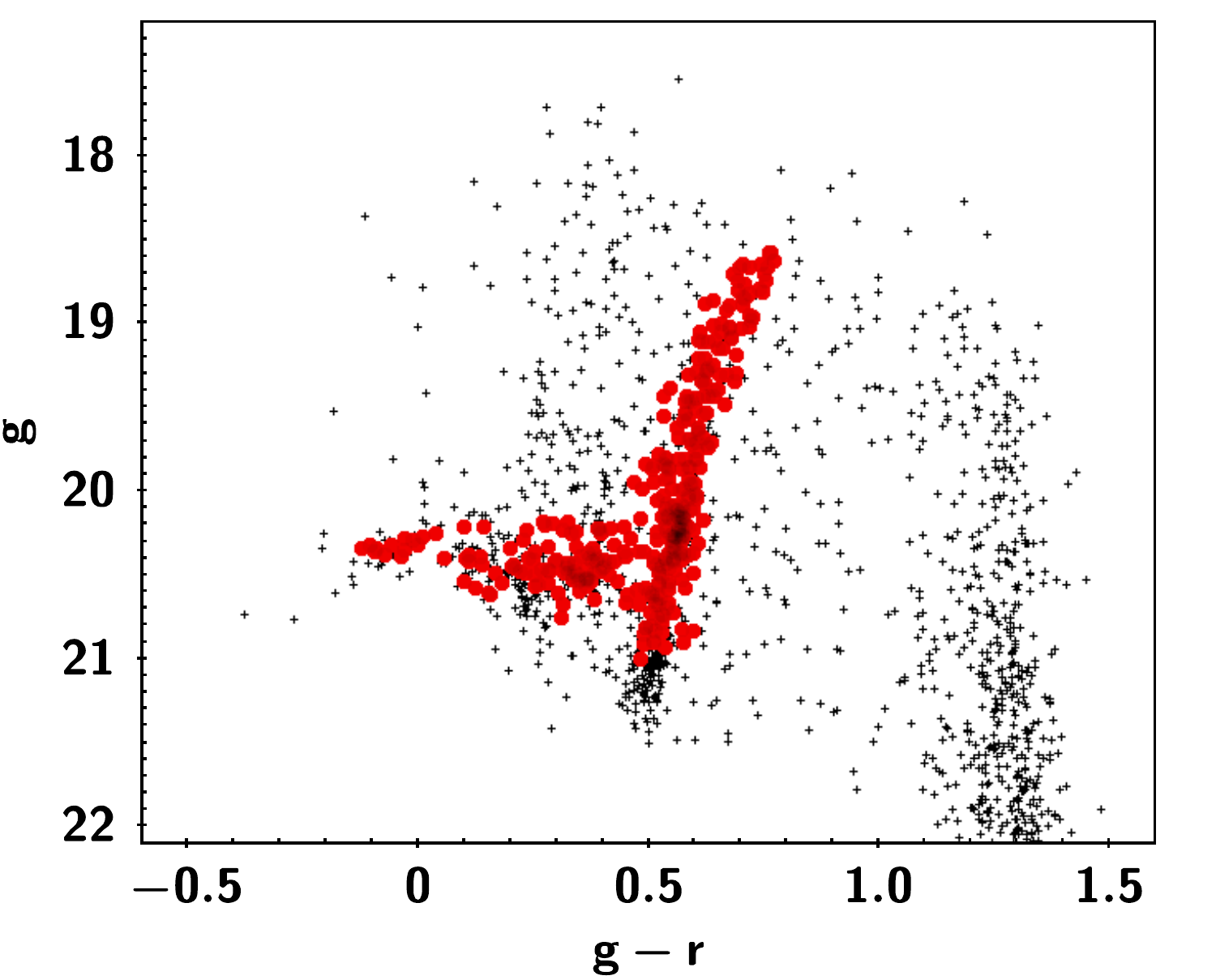}
\includegraphics[width=0.45\textwidth]{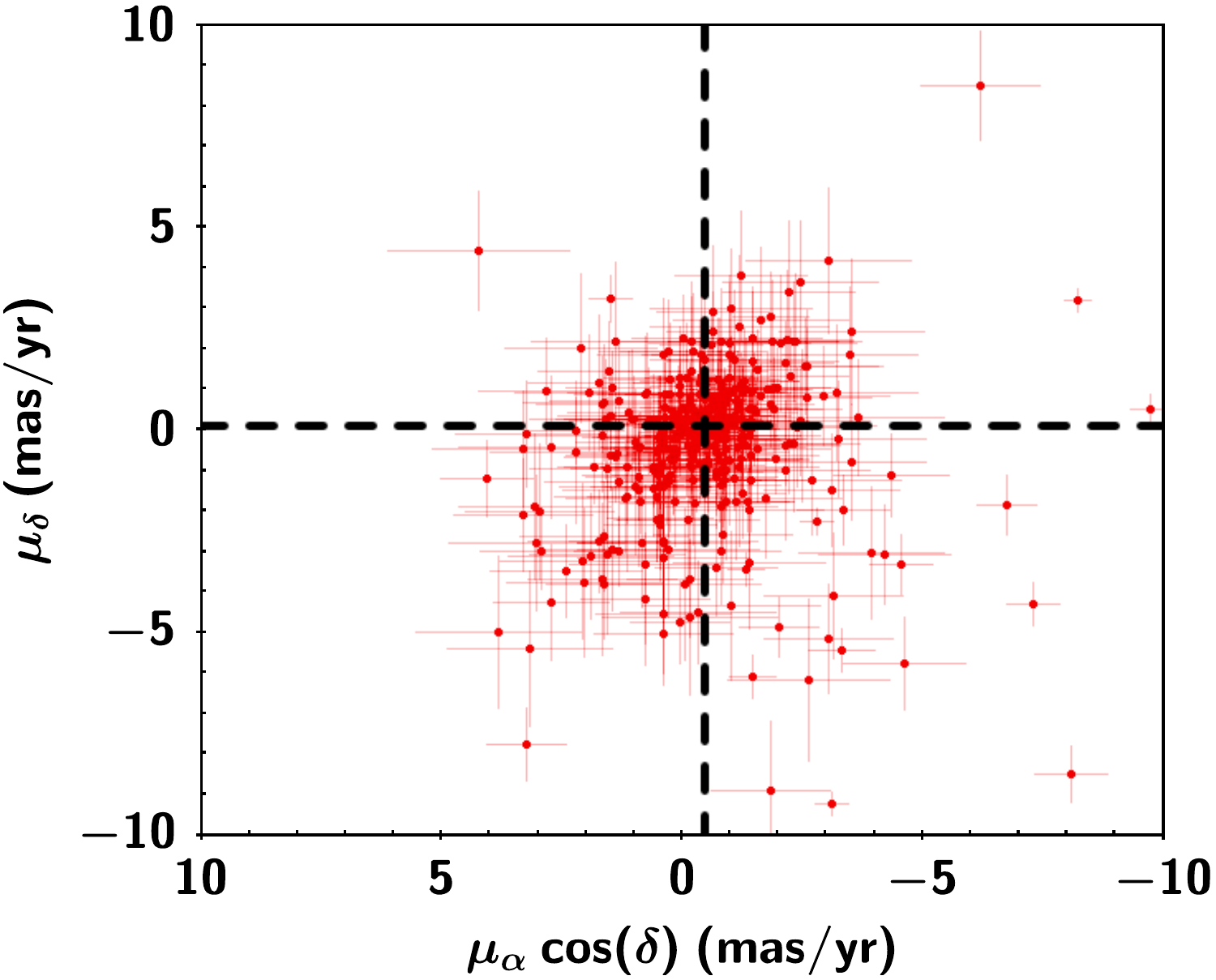}
\includegraphics[width=0.45\textwidth]{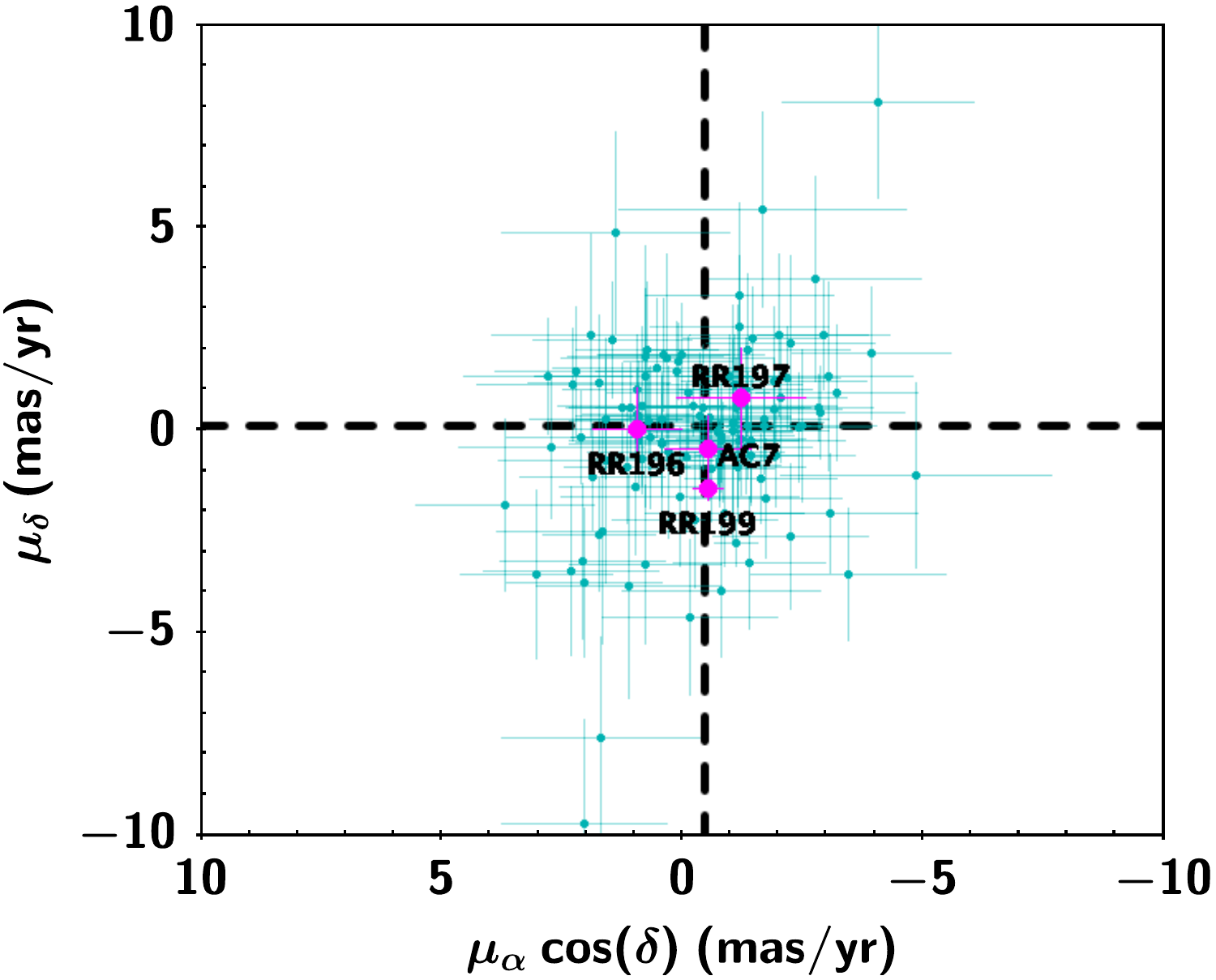}
\includegraphics[width=0.45\textwidth]{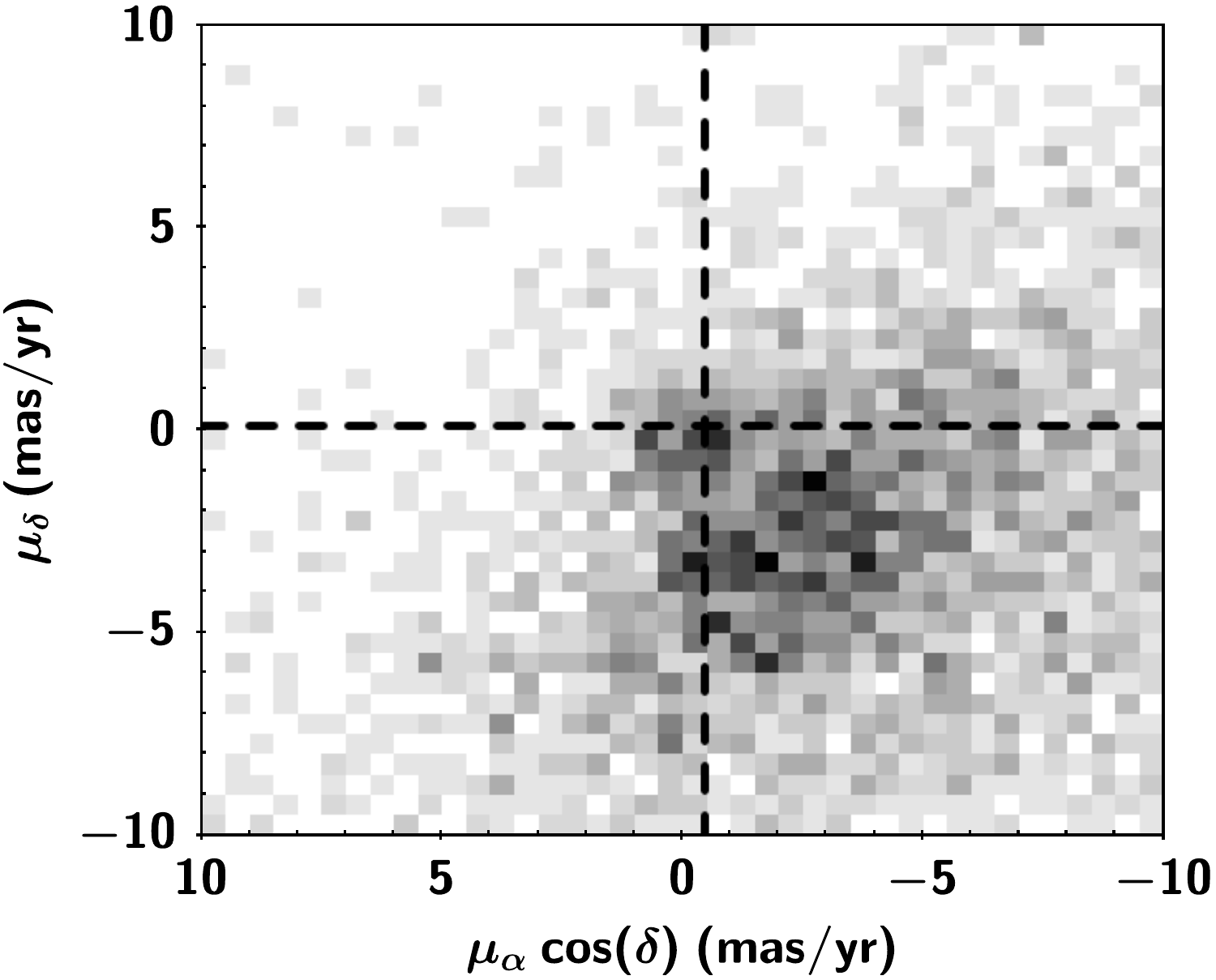}
\caption{(Top Left) CMD of Gaia DR2 stars that matched with our catalog in the region within the core radius of Sextans. We selected from there only stars with error in the proper motions $<2$ mas/yr and in either the red giant branch or horizontal branch of Sextans (red circles). (Top Right) Proper motion distribution of Sextans stars as selected in the previous panel. (Bottom Left) Proper motion of 117 variable stars in our survey that matched Gaia DR2. The extra-tidal candidates have been labeled. (Bottom, Right) Proper motion of foreground population. Only stars with proper motion errors $<1.0$ mas/yr were included in this panel.}
\label{fig:gaia}
\end{figure*}

Figure~\ref{fig:sky-var} shows that the distribution in the sky of the bulk of the variables detected in this work is concentrated toward the center of the galaxy, in a rather "boxy" shape following the distribution of the Sextans stellar population \citep[see stellar density maps in Fig 7 and 13 in][]{roderick16}. More than half of the variable stars  (58\%) are indeed contained within the $26\farcm 8$ core radius of Sextans. This central concentration distribution holds for the three type of pulsating variables that we have found in this work. This is not a selection effect since, actually, we have better temporal coverage and deeper observations in the two outermost fields of our survey.

There are 3 variable stars, 2 RR Lyrae stars (RR196 and RR197) and 1 Anomalous Cepheid (AC7) which lie outside the $83\farcm 2$ tidal radius of Sextans. As discussed before (\S\ref{sec:obs}), the above tidal radius \citep{roderick16} is significantly smaller than the ones given in other works in the literature. In the discovery paper, \citet{irwin90} measured a tidal radius of $90\arcmin$, which was revisited later by \citet{irwin95} to $r_t=160\arcmin$. Recently, both \citet{okamoto17} and \citet{cicuendez18} suggested also a large tidal radius, $120\arcmin$. Thus, the possibility of being extra-tidal stars only holds if the tidal radius is indeed as short as suggested by \citet{roderick16}. 

The three stars presumed to be extra-tidal are located toward the NE side of Sextans. This, however, may be a selection effect since we covered the galaxy only along the semi-major axis and thus, we cannot say if there is debris along the semi-minor axis.  Evidence for extra-tidal material in this galaxy is hard to prove because of its large low surface brightness and large field contamination specially at large distances from its center. Variable stars are then a particular useful tracer of possible extra-tidal material and the stars we have found may indicate the galaxy is disrupting. In order to further test this scenario we studied the proper motions of the variable stars in Sextans in Gaia DR2 \citep{gaia16,gaia18}. First, we isolated Sextans stars (variable or not) by selecting from all DR2 objects within the core radius of Sextans (black $+$ symbols in the upper left panel of Figure~\ref{fig:gaia}) those in the red giant branch and horizontal branch of Sextans that have errors in the proper motion $<2$ mas/yr (red circles in Figure~\ref{fig:gaia}, top, left). The proper motions of that sample of $\sim 1,000$ Sextans stars are shown in the top right panel of Figure~\ref{fig:gaia}. The vertical and horizontal dashed lines indicate the mean proper motion, $(\mu_\alpha*,\mu_\delta) = (-0.496, 0.077)$ mas/yr, as derived by \citet{helmi18}. Our selection of Sextans stars are clumped together in proper motion space around those values. For comparison, we show in the bottom right panel the proper motion distribution of non-Sextans stars in the field. To select this population we used our full catalog but did not include either stars inside the locus of the HB and red giant branch nor stars with proper motion errors $>1.0$ mas/yr.  Although some contamination by Sextans stars seems to still be present in this figure, it is clear that the foreground population in the field concentrates in the fourth quadrant of this diagram (negative proper motions in both coordinates) which mostly separates from the bulk of the Sextans population. 

We then matched our list of variable stars with Gaia DR2. Out of our list of 222 variables stars (of all types), a total of 117 have measured proper motions in the Gaia DR2 dataset. As expected, none of the DC stars in our sample had a counterpart in DR2 since these stars are beyond the Gaia limiting magnitude. All the Anomalous Cepheids but AC1 had a proper motion measurement. In Figure~\ref{fig:gaia} (right) we show the proper motion of the RR Lyrae stars and Anomalous Cepheids found in Gaia. As expected, they nicely clump together around the Sextans proper motions. Our 3 possible extra-tidal stars are labeled and shown with large symbols in Figure~\ref{fig:gaia} (right). All three of them have proper motions in agreement with being Sextans members, and may be extra-tidal material. In particular, AC7, which is a bright star ($g=19.72$), has relatively small error bars in its proper motion ($\sigma\mu_\alpha = 0.90$ mas/yr, $\sigma\mu_\delta = 0.83$ mas/yr) and it is within $1\sigma$ from the mean proper motion of Sextans. 

In the right panel of Figure~\ref{fig:gaia}  we also highlight the case of RR199 which is one of the two RR Lyrae stars that we classified as a field variable. The other star is RR198 which unfortunately has no counterpart in Gaia DR2.  These two stars were also located outside the tidal radius of Sextans (blue squares in Figure~\ref{fig:sky-var}) but were marked as RR Lyrae stars because of their location as well as the fact that they were quite bright, brighter than the rest of the Anomalous Cepheid stars. The period range of Anomalous Cepheids overlaps with that of the RR Lyrae stars and there is no easy way to distinguish one from the other if the distance is not known \citep{catelan15}. If they were RR Lyrae stars, their magnitudes locate these stars at 28 and 29 kpc from the Sun. We expect 1 ($\pm1$) halo RR Lyrae star in our footprint with distances between 25 and 35 kpc, according to the number density profile of the halo by \citet{medina18}.  The proper motion of RR199 however shows compatibility with being a Sextans member. This opens the possibility that RR199 is not really a field RR Lyrae star but a extra-tidal Anomalous Cepheid in Sextans, similar in that respect to AC7.  RR199 is separated by 2 mags from the mean of the RR Lyrae stars in the Sextans HB (open blue squares in Figure~\ref{fig:CMDvar}). This large $\Delta {\rm mag}$ between the Anomalous Cepheid and the HB has been observed also in other systems, for example, in Carina \citep{coppola15} and Leo I \citep{stetson14}, although the latter may contain a mix of Anomalous Cepheids and short-period Classical Cepheids. For RR198 there are no proper motions available but it has similar properties to RR199 in the CMD. Both stars are also located toward the NE side of Sextans, similar to the other extra-tidal variable stars mentioned above. Radial velocities are needed to settle if they are indeed MW foreground RR Lyrae stars or Sextans Anomalous Cepheid variables.
 
\section{The Distance to Sextans from its RR Lyrae Stars} \label{sec:distance}

\begin{figure}[htb!]
\epsscale{1.2}
\plotone{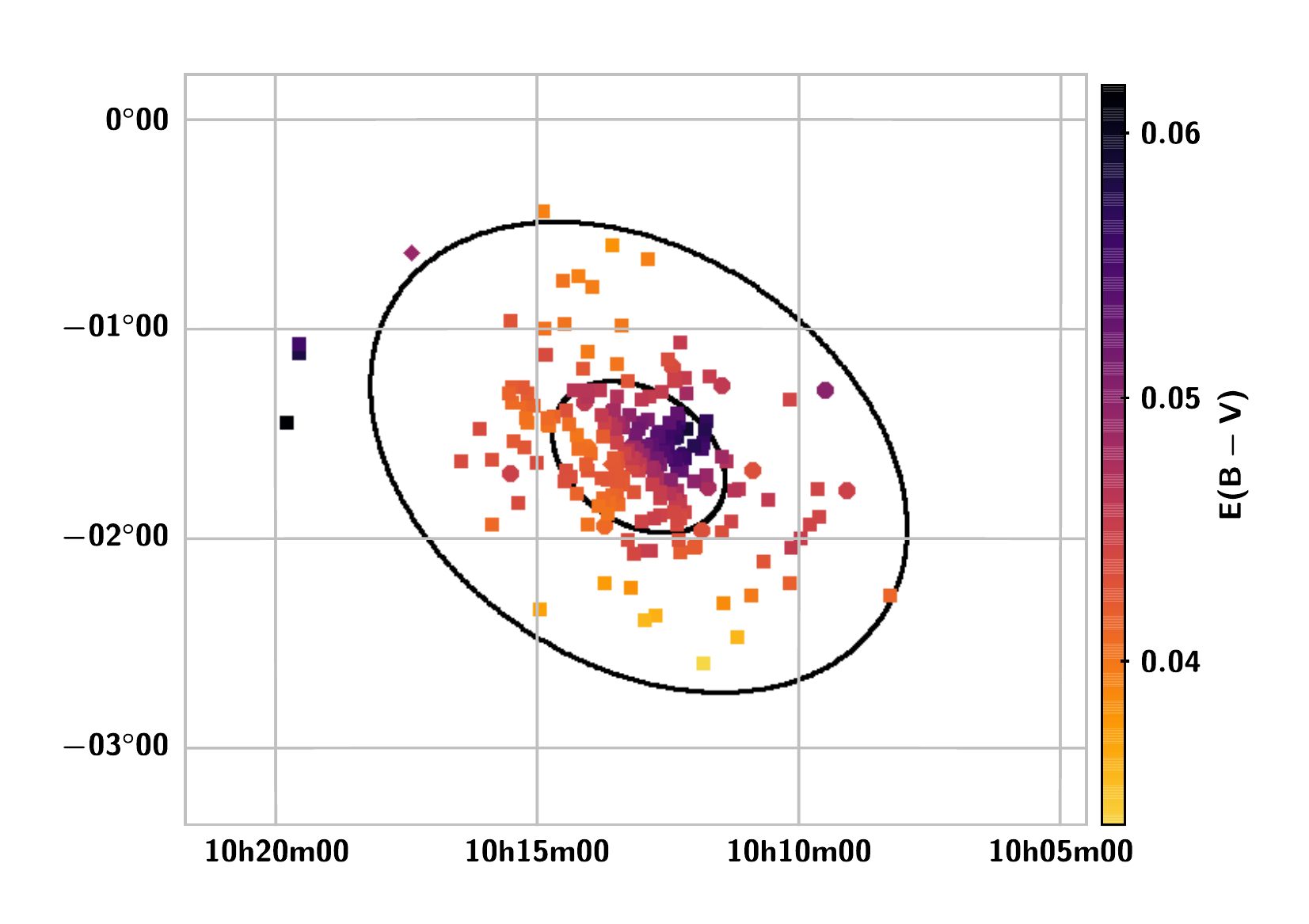}
\caption{Map (RA versus DEC) of the color excess E(B-V) from \citet{schlegel98} toward each variable star found in this work (squares, RR Lyrae stars; circles, dwarf Cepheids; diamonds, Anomalous Cepheids). The two ellipses correspond to the $r_c$ and $r_t$ as determined by \citet{roderick16}.}
\label{fig:ext}
\end{figure}

\begin{figure}[htb!]
\centering
\includegraphics[width=9.0cm]{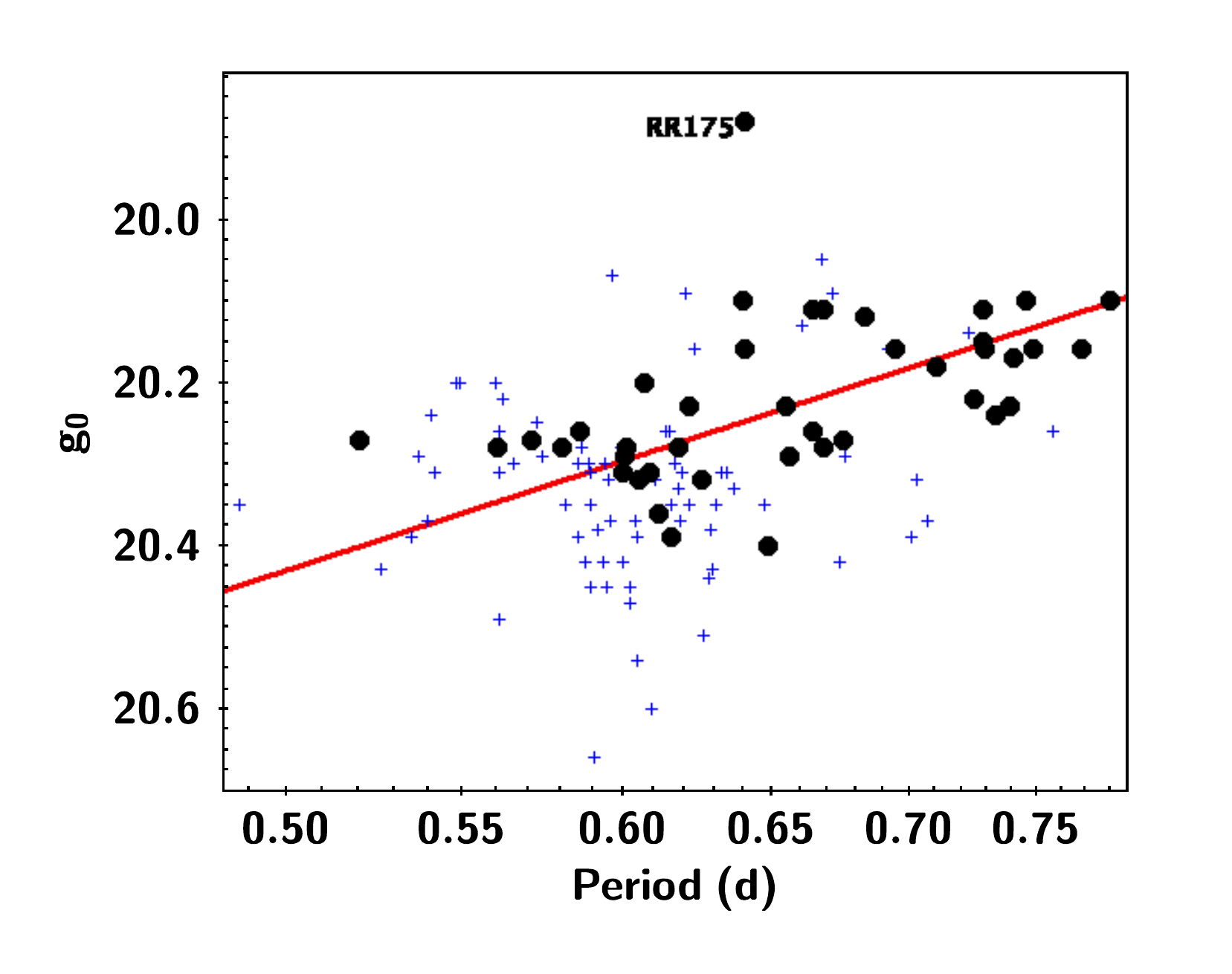}
\includegraphics[width=9.0cm]{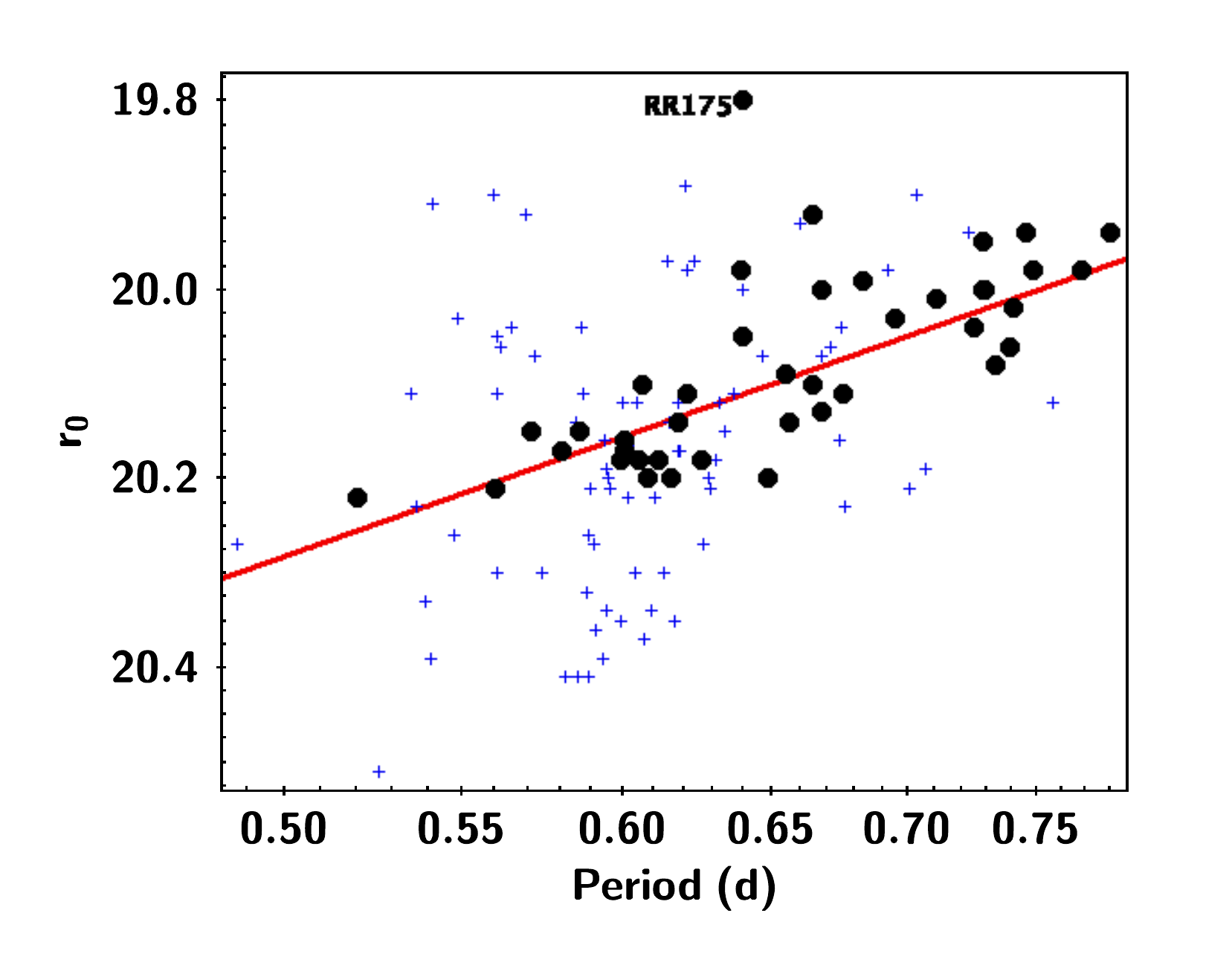}
\caption{Extinction corrected magnitudes $g_0$ (Top) and $r_0$ (Bottom) of the Sextans RR Lyrae stars versus period. RR Lyrae stars in Fields A and B which have complete lightcurves available are shown with large solid circles. Stars in Field C, which have only partially observed light curves are shown with $+$ symbols. The red line is not a fit to our data but a representation of equations~\ref{eq:PLZg} and ~\ref{eq:PLZr} with [Fe/H]$=-1.93$ and shifted by a distance modulus of 19.64 and 19.68 in $g$ and $r$ respectively. RR175 may be an evolved RR Lyrae star or an Anomalous Cepheid. 
}
\label{fig:PLZ_RRLS}
\end{figure}

Before introducing the behavior of the Sextans DC stars in a PL diagram and comparing with other works, we will anchor the distance of Sextans to its RR Lyrae stars. Mean magnitudes for the variable stars were corrected by interstellar extinction using the following equations with coefficients taken from \citet{schlafly11} for PS1 magnitudes (their Table 6), in combination with the color excesses $E(B-V)$ from the dust maps of \citet{schlegel98}: $A_g = 3.172 E(B-V)$ and $A_r = 2.271 E(B-V)$. A map of the color excess in the Sextans region is shown in Figure~\ref{fig:ext}. Although this is a high latitude region, the area covered by Sextans is large enough that significant differences in reddening are observed throughout the galaxy.  

We used the following $g$ and $r$ Period-Luminosity-Metallicity (PLZ) relationship for {\rrab } stars in the PS1 magnitude system, taken from \citet{sesar17}:

\begin{equation}
M_g = -1.7  \log(P/P_{\rm ref}) + 0.08 (\rm{[Fe/H]} - \rm{[Fe/H]_{ref}} ) + 0.69
\label{eq:PLZg}
\end{equation}

\begin{equation}
M_r = -1.6  \log(P/P_{\rm ref}) + 0.09 (\rm{[Fe/H]} - \rm{[Fe/H]_{ref}} ) + 0.51
\label{eq:PLZr}
\end{equation}

\noindent 
where $P_{\rm ref} = 0.6$d and [Fe/H]$_{\rm ref}= -1.5$ dex. The rms of these relationships are 0.07 and 0.06 mags in $g$ and $r$, respectively. We applied these PLZ
relationships assuming a mean metallicity for Sextans of [Fe/H]$-1.93$ dex \citep{kirby11}. The dependence with period of these relationships is in excellent agreement with our data (Figure~\ref{fig:PLZ_RRLS}). To calculate the mean distance modulus to Sextans from the {\rrab} stars we used only the 42 stars for which we had complete coverage of their lightcurves
(the ones in fields A and B). They are shown as solid circles in Figure~\ref{fig:PLZ_RRLS}. Although we fitted reasonably well templates to the $g$ light curves of stars in Field C, for many stars the coverage of the observations is not enough to cover the full amplitude of the variation, and there is uncertainty in the final amplitudes. This is particularly true for stars which do not have observations near maximum light (see for example RR29 in Figure~\ref{fig:lcRR}). Consequently, the mean magnitude of the RR Lyrae stars in field C, which is calculated integrating the fitted template, may not be accurate. Figure~\ref{fig:PLZ_RRLS} shows the stars in Field C ("$+$" symbols in Figure~\ref{fig:PLZ_RRLS}) also follow the PLZ of the well observed stars but, understandably, they display more dispersion. In the $r$ band the dispersion is even larger since there are only 4 observations per star in this band and the mean magnitudes are straight averages of those. Among the sample of well observed RR Lyrae stars, one star, RR175, is significantly brighter than the rest of the stars in both bands. It is possible that RR175 is either an evolved RR Lyrae star or an Anomalous Cepheid. 

The mean distance modulus and error of the mean from the $g$ lightcurves of 41 {\rrab } stars (excluding RR175) are $19.64\pm 0.01$ mag, while in the $r$ band, we obtained $19.677\pm 0.008$. These are equivalent to 84.7 and 86.2 kpc from the Sun, which are in excellent agreement with previous estimates, for example,  84.2 kpc from \citet{medina18} and 86 kpc from \citet{mateo95} and \citet{lee09}. Because most of the DC stars were observed only in the $g$ band, we will assume the distance modulus from that band in the following analysis.

\section{The Period-Luminosity relationship of DC stars in Sextans} \label{sec:PL-DC}

\begin{figure}[htb!]
\epsscale{1.2}
\plotone{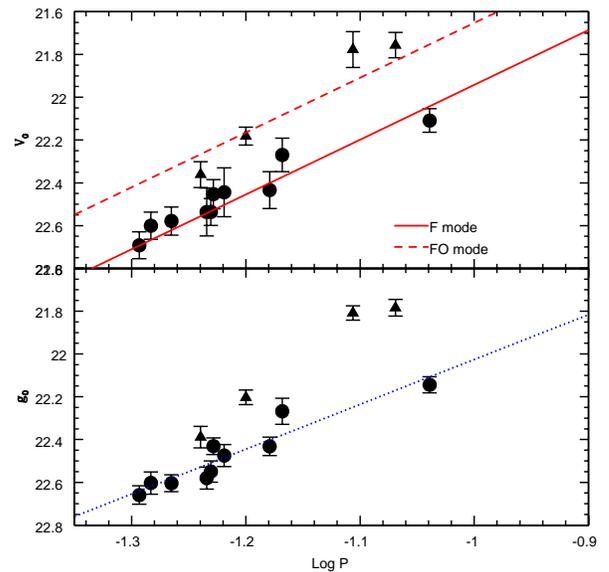}
\caption{(Top) Extinction corrected V magnitude (transformed from our $g$ and $r$) versus the logarithm of the pulsation period for the 14 DC stars found in Sextans. The red lines correspond to the PLZ relationships given by \citep{nemec94} for F and FO pulsators, shifted by the distance modulus found with the RR Lyrae stars, $\mu_0=19.64$ and assuming [Fe/H]$_{\rm ref}= -1.93$ dex. 
Based on these relationships, the stars identified with triangles were associated with the FO mode. (Bottom) Extinction corrected $g$ magnitude versus the logarithm of the pulsation period. The dotted blue line is a fit to the alleged F-mode pulsating stars (see text).}
\label{fig:PL_DC}
\end{figure}

A plot of the extinction-corrected $g$ magnitude, $g_0$, versus the logarithm of the period (lower panel in Figure~\ref{fig:PL_DC}) of the 14 DC stars in Sextans shows a clear trend with the shorter period variables being fainter than the ones with longer period. There is a lot of scatter in the data though, likely due to the fact that our sample may contain stars pulsating in different modes. As discussed before (\S\ref{sec:DC}) there is no unambiguous way to recognize a fundamental (F) mode pulsator from a first overtone (FO) one with the data at hand, and different pulsation modes obey different PLZ relationships \citep[e.g.][]{nemec94}. 

Since the distance to Sextans has been set by the RR Lyrae stars, we can use known PLZ relationships of DC stars to infer their pulsation mode. \citet{nemec94} derived such relationships but in the Johnson V band. Thus, we transformed our mean $g$ and $r$ magnitudes (in the PS1 system) of the DC stars in Sextans to V using:

\begin{equation}
V=g+0.006-0.525*(g-r)
\end{equation}

\noindent
from \citet{tonry12}. The extinction corrected $V_0$ magnitudes \citep[obtained using][]{schlafly11} versus the logarithm period are shown in the top panel of Figure~\ref{fig:PL_DC}, which also shows both the F and FO relationships from \citet{nemec94} shifted to a distance modulus of 19.64 mag (from the RR Lyrae stars), and assuming [Fe/H]$= -1.93$ dex. The agreement of those relationships with our data is very good, and we can infer from there that there are 4 stars pulsating in the FO mode. These stars are indicated with triangles in Figure~\ref{fig:PL_DC}. Assuming that the remaining 10 stars are then F pulsators, we obtain, via least square fitting, the following relationships for Sextans:

\begin{equation}
V_0= -2.22 \log P + 19.77; \;\;\; rms=0.04
\end{equation}

\begin{equation}
g_0 = -2.10 \log P + 19.93; \;\;\; rms=0.05
\end{equation}

Anchoring these equations to the distance modulus given by the RR Lyrae stars,  the PL relationship for Sextans (F mode) DC stars are:

\begin{equation}
M_V = -2.10 \log P + 0.13
\end{equation}

\begin{equation}
M_g = -2.01 \log P + 0.29
\end{equation}

The resulting slope in our $V$-PL relationship is, within errors, compatible with the one provided by \citet[][$-2.56\pm0.54$]{nemec94}. Surprisingly, the slope is somewhat shallower than the slopes given by \citet{mcnamara11,cohen12,poretti08} and \citet{fiorentino15}. The slope of the PL in Carina \citep{vivas13}, however, is even shallower ($-1.68$). These apparent discrepancies between different works demonstrate the need to continue gathering data of DC stars in other systems in order to study the effects of local conditions (age, metallicity, star formation history) in the PL relationship.

\section{The DC population in Sextans and other galaxies} \label{sec:propertiesDC}

To date, DC stars have been searched and found in 6 of the ''classical'' satellites of the Milky Way, including this work, and in 2 Ultra-Faint Dwarf (UFD) satellites.  The number and mean period of the DC population in those systems is summarized in Table~\ref{tab:allDC}.
The search in Sagittarius, however, is incomplete since no dedicated survey in this galaxy has been done. The 3 stars listed in Table~\ref{tab:allDC} refers to stars serendipitously discovered in the background of the M55 globular cluster.

\begin{deluxetable*}{lrcr}[htb!]
\tablecolumns{4}
\tablewidth{0pc}
\tablecaption{Known Dwarf Cepheid Stars in Extragalactic Systems \label{tab:allDC}}
\tablehead{
Galaxy & $N_{DC}$ & Mean Period (d) & Reference \\
}
\startdata
\cutinhead{Classical Satellites}
LMC      &  2,323   &   0.074  &  \citet{garg10} (SuperMACHO) \\
             &  1,276  &   0.110  &   \citet{poleski10} (OGLE-III)\tablenotemark{a} \\
Fornax   & 85        &  0.070  & \citet{poretti08} \\
Carina   &     426   &   0.060  &   \citet{vivas13,coppola15}\tablenotemark{b} \\
Sculptor   &   23     &  0.066   & \citet{martinez16}  \\
Sagittarius & 3 & 0.0481 & \citet{pych01}\tablenotemark{c} \\
Sextans  &    14    &  0.065   & This Work   \\
\cutinhead{UFD Satellites}
Leo IV  &  1 & 0.099 & \citet{moretti09} \\
Coma Berenices & 1 & 0.125 & \citet{musella09}
\enddata
\tablenotetext{a}{The original OGLE-III catalog contains 2,786 stars but we are not including here either stars flagged as {\sl uncertain} nor Galactic stars based on proper motions \citep{poleski10}.}
\tablenotetext{b}{Both catalogs were merged and duplicates were eliminated.}
\tablenotemark{c}{In the background of the M55 globular cluster.}
\end{deluxetable*} 

\begin{figure}[htb!]
\centering
\includegraphics[width=9cm]{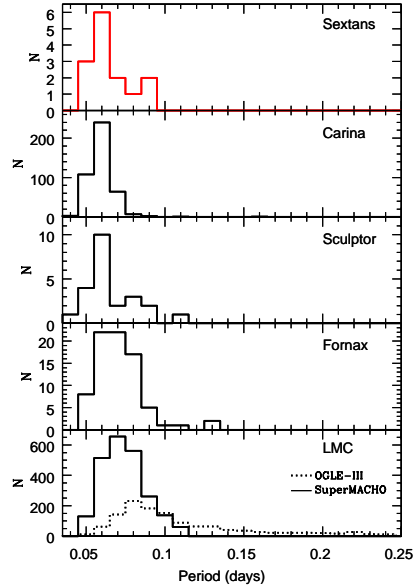}
\caption{Period distribution of  DC stars in the 5 classical satellites with known DC stars. References for each galaxy are given in Table~\ref{tab:allDC}}.
\label{fig:per_DC}
\end{figure}

In Figure~\ref{fig:per_DC} we show the period distribution of the DC stars in the 5 satellite galaxies with a large number of known DC stars. We do not include either Leo IV nor Coma Berenices because they have only 1 star each (see Table~\ref{tab:allDC}). The figure includes DC stars in the LMC from two different experiments: OGLE-III \citep{poleski10} and Super-MACHO \citep{garg10}. The period distribution is quite different among those two samples that we show them separately. The DC stars in the SuperMACHO sample \citep{garg10} have a mean period of 0.074 d and the distribution is not that different from that of Fornax, which has a mean period of 0.07 days. On the other hand, \citet{poleski10} shows a much wider period distribution among the DC stars measured by OGLE-III, with a long tail toward long periods, which is not observed in any of the other data-sets. The DC stars in Sextans have a very similar mean period as the ones in Sculptor \citep{martinez16}. As Sculptor, the distribution shows a peak at $\sim 0.055$ d but have a tail toward longer periods. None of these two galaxies have stars with periods longer than 0.12 d. Carina's sample also peaks at the same period but the distribution is more symmetrical with no extension toward longer periods.  Surprisingly, the 2 DC stars in the UFD galaxies have significant longer periods (0.099 and 0.125 d), than the mean of the other galaxies. To search for DC stars in other UFDs will be interesting in order to confirm this trend.

The Sextans population of variables is not that different from Sculptor. Sextans has 14 DC stars and 227 RR Lyrae stars. On the other hand, from \citet{martinez16}, Sculptor has 23 DC stars and  523 RR Lyrae stars. The ratio of DC stars to RR Lyrae stars is  then 1:16 and 1:22 in Sextans and Sculptor, respectively. On the contrary, the ratio is $\sim$1:1 in Fornax \citep{poretti08}, and  Carina stands significantly apart from these other galaxies in having a ratio of 5:1, meaning it contains more DC than RR stars. Both Sextans and Sculptor  are galaxies that are dominated by an old population with only small contributions from younger stellar populations, which is not the case for the other classical dwarfs in Table~\ref{tab:allDC}.

\section{Summary and Conclusions} \label{sec:conclu}

A full survey of Sextans had been difficult in the past due to its large extension in the sky and large distance from the Sun. The combination of the large FoV of DECam with a 4m telescope allowed us  to search for variable stars in the instability strip of the Sextans dSph galaxy, down to magnitude $g\sim 23$. The survey, covering 7 sq degrees, covers a large part of the galaxy. Although the true size of Sextans is controversial, if the small scale tidal radius by \citet{roderick16} is confirmed, this would be the first time that a full census of pulsating variable stars is done in the whole galaxy. We found 7 Anomalous Cepheids, 197 RR Lyrae stars and 14 DC. In addition, there are 4 stars with properties of either RR Lyrae stars or DC but in a location within  the CMD that suggest they are rather Milky Way foreground stars. In this paper we focused on two aspects of the survey: {\sl (i)} the spatial distribution of the variable stars, and {\sl (ii)} the properties of the DC population. 

Variable stars have been used in the past to trace extra-tidal material in stellar systems. For example, one of the first hints on the existence of long tidal tails coming from the Sagittarius dwarf came indeed from RR Lyrae stars \citep{mateo96,vivas01}. Extra tidal variable stars have also been found in Carina \citep{vivas13}, Hercules \citep{garling18} and Tucana III (Mart{\'\i}nez-V\'azquez et al, in preparation). Our survey extends beyond the \citet{roderick16}'s tidal radius of Sextans along the direction of the semi-major axis and provides an opportunity to trace any possible sign of debris outside the tidal radius. Indeed, we found 2 RR Lyrae stars and 1 Anomalous Cepheid outside the tidal radius of Sextans. Not only these stars are located in the right place in the CMD to be considered Sextans members, but also they have Gaia proper motions consistent with the bulk of the Sextans population. There is an additional extra-tidal RR Lyrae star in La Silla-QUEST survey which is just outside our footprint. A very rough estimate of the amount of tidal debris can be obtained considering there are 3 extra tidal RR Lyrae stars out of a total of 227 (including the ones from other surveys that we missed due to the CCD gaps). This suggest a minimum of 1\% of material outside the tidal radius. That is just a lower limit because our survey do not cover regions outside Sextans along the semi-minor axis and thus, if the extra-tidal material is more uniformly distributed, we can be missing part of it. If Sextans is larger as suggested by \citet{irwin95}, \citet{okamoto17} and \citet{cicuendez18}, all variable stars are inside the tidal radius. In that case, our survey does not have enough coverage to study if extra-tidal material indeed exists.

In an analysis of the stellar populations of Sextans done, also with DECam, by \citet{roderick16}, the authors conclude that the Blue Straggler (BS) population is more centrally concentrated that the Blue HB population which suggest that most of the BS stars may be main sequence stars from an intermediate-age population. The specific star formation history of Sextans may be responsible of  such population gradient \citep{lee09}. We find support to this scenario with our variable stars. The RR Lyrae stars, an unequivocal old-age tracer, are more widely distributed than the DC population, which are variable BS stars. No DC stars were found in the external part of Sextans. An interesting additional puzzle comes from the Anomalous Cepheid stars. There is, at least, 1 extra tidal Anomalous Cepheid in Sextans, but there is a possibility that two stars that we classify originally as foreground RR Lyrae stars are indeed Anomalous Cepheid, in which case, they would also be extra-tidal stars. Anomalous Cepheids are usually regarded as intermediate-age population but they do not have the same spatial distribution as the DC stars. A possible explanation to this apparent discrepancy may come from the idea that some Anomalous Cepheids are produced through binary evolution \citep{gautschy17}, of old stars in this case.

The second aspect in which we focused in this paper was the DC population. With the DC stars found in Sextans, there are now 5 local group galaxies which have been adequately surveyed for this type of stars: LMC, Fornax, Carina, Sculptor and Sextans. These galaxies cover a wide range of properties and different star formation histories. The DC population seems to change accordingly. Sextans and Sculptor, two galaxies dominated by an old population, have relatively few DC stars. The RR Lyrae variables dominate the number of pulsating variables in those galaxies. On the contrary, galaxies with  strong young and intermediate age population have a rich population of DC stars; in the case of Carina, for example, there are several times more DC than RR Lyrae stars. The increased census of known DC stars in these different systems provide constraints on models of production of DC stars under different environmental conditions.

The increased census is also useful to investigate the use and limitations of DC stars as standard candles. In this work we set a PL relationship which is anchored to the distance modulus obtained by the RR Lyrae stars. The resulting PL relationship has a dispersion of only 0.04 mag. In the future, we plan to explore the possibility that a metallicity-independent PL relationship like the one obtained by \citet{cohen12} with data from only a few galaxies, can be used globally. 

Sextans has a rich population of RR Lyrae stars compared with its DC population. It is clear then that RR Lyrae stars would be the ideal choice for standard candles in this galaxy. They are not only more numerous but also brighter than DC stars. However, not necessarily RR Lyrae stars are as abundant in other systems. Carina 
\citep{vivas13,coppola15} has more DC stars than RR Lyrae stars. Leo T \citep{clementini12} has only 1 RR Lyrae stars but several Anomalous Cepheids. Just recently, a very young, disrupting open cluster was discovered in the Milky Way's halo \citep{price-whelan18}; given its young age, this system presumably does not have RR Lyrae stars. These examples illustrate the importance of setting up alternative standard candles that trace different stellar populations. DC stars will be observed by LSST to large distances and will potentially provide a tool to study halo substructures and stellar populations to very large distances in the Galactic halo. 

\bigskip
\bigskip

\acknowledgments

We thank the anonymous referee for an insightful review of this manuscript.
We thank Marcio Catelan and P{\'\i}a Amigo for providing us with Amigo (2012)'s PhD thesis manuscript which contains the full
table of variable stars in Sextans detected in her work. J.A-G. acknowledges support by FONDECYT Iniciación 11150916 and by the Ministry of Economy, Development, and Tourism’s Millennium Science Initiative through grant IC120009, awarded to the Millennium Institute of Astrophysics (MAS). MM acknowledges support from NSF grants AST-1312997 and AST-1815403.

This project used data obtained with the Dark Energy Camera (DECam),
which was constructed by the Dark Energy Survey (DES) collaboration.
Funding for the DES Projects has been provided by 
the U.S. Department of Energy, 
the U.S. National Science Foundation, 
the Ministry of Science and Education of Spain, 
the Science and Technology Facilities Council of the United Kingdom, 
the Higher Education Funding Council for England, 
the National Center for Supercomputing Applications at the University of Illinois at Urbana-Champaign, 
the Kavli Institute of Cosmological Physics at the University of Chicago, 
the Center for Cosmology and Astro-Particle Physics at the Ohio State University, 
the Mitchell Institute for Fundamental Physics and Astronomy at Texas A\&M University, 
Financiadora de Estudos e Projetos, Funda{\c c}{\~a}o Carlos Chagas Filho de Amparo {\`a} Pesquisa do Estado do Rio de Janeiro, 
Conselho Nacional de Desenvolvimento Cient{\'i}fico e Tecnol{\'o}gico and the Minist{\'e}rio da Ci{\^e}ncia, Tecnologia e Inovac{\~a}o, 
the Deutsche Forschungsgemeinschaft, 
and the Collaborating Institutions in the Dark Energy Survey. 
The Collaborating Institutions are 
Argonne National Laboratory, 
the University of California at Santa Cruz, 
the University of Cambridge, 
Centro de Investigaciones En{\'e}rgeticas, Medioambientales y Tecnol{\'o}gicas-Madrid, 
the University of Chicago, 
University College London, 
the DES-Brazil Consortium, 
the University of Edinburgh, 
the Eidgen{\"o}ssische Technische Hoch\-schule (ETH) Z{\"u}rich, 
Fermi National Accelerator Laboratory, 
the University of Illinois at Urbana-Champaign, 
the Institut de Ci{\`e}ncies de l'Espai (IEEC/CSIC), 
the Institut de F{\'i}sica d'Altes Energies, 
Lawrence Berkeley National Laboratory, 
the Ludwig-Maximilians Universit{\"a}t M{\"u}nchen and the associated Excellence Cluster Universe, 
the University of Michigan, 
{the} National Optical Astronomy Observatory, 
the University of Nottingham, 
the Ohio State University, 
the OzDES Membership Consortium
the University of Pennsylvania, 
the University of Portsmouth, 
SLAC National Accelerator Laboratory, 
Stanford University, 
the University of Sussex, 
and Texas A\&M University.

Based on observations at Cerro Tololo Inter-American Observatory, National Optical
Astronomy Observatory (NOAO Prop. IDs: 2014A-0313, 2017A-0951, PI: Vivas), which is operated by the Association of
Universities for Research in Astronomy (AURA) under a cooperative agreement with the
National Science Foundation.

The Pan-STARRS1 Surveys (PS1) and the PS1 public science archive have been made possible through contributions by the Institute for Astronomy, the University of Hawaii, the Pan-STARRS Project Office, the Max-Planck Society and its participating institutes, the Max Planck Institute for Astronomy, Heidelberg and the Max Planck Institute for Extraterrestrial Physics, Garching, The Johns Hopkins University, Durham University, the University of Edinburgh, the Queen's University Belfast, the Harvard-Smithsonian Center for Astrophysics, the Las Cumbres Observatory Global Telescope Network Incorporated, the National Central University of Taiwan, the Space Telescope Science Institute, the National Aeronautics and Space Administration under Grant No. NNX08AR22G issued through the Planetary Science Division of the NASA Science Mission Directorate, the National Science Foundation Grant No. AST-1238877, the University of Maryland, Eotvos Lorand University (ELTE), the Los Alamos National Laboratory, and the Gordon and Betty Moore Foundation.

This work has made use of data from the European Space Agency (ESA) mission
{\it Gaia} (\url{https://www.cosmos.esa.int/gaia}), processed by the {\it Gaia}
Data Processing and Analysis Consortium (DPAC,
\url{https://www.cosmos.esa.int/web/gaia/dpac/consortium}). Funding for the DPAC
has been provided by national institutions, in particular the institutions
participating in the {\it Gaia} Multilateral Agreement.

\vspace{5mm}
\facility{Blanco (DECam)}


\bibliographystyle{aasjournal}
\bibliography{Sextans}

\begin{thebibliography}{}
\expandafter\ifx\csname natexlab\endcsname\relax\def\natexlab#1{#1}\fi
\providecommand{\url}[1]{\href{#1}{#1}}
\providecommand{\dodoi}[1]{doi:~\href{http://doi.org/#1}{\nolinkurl{#1}}}
\providecommand{\doeprint}[1]{\href{http://ascl.net/#1}{\nolinkurl{http://ascl.net/#1}}}
\providecommand{\doarXiv}[1]{\href{https://arxiv.org/abs/#1}{\nolinkurl{https://arxiv.org/abs/#1}}}

\bibitem[{{Alonso-Garc{\'{\i}}a} {et~al.}(2012){Alonso-Garc{\'{\i}}a}, {Mateo},
  {Sen}, {Banerjee}, {Catelan}, {Minniti}, \& {von Braun}}]{alonso12}
{Alonso-Garc{\'{\i}}a}, J., {Mateo}, M., {Sen}, B., {et~al.} 2012, \aj, 143,
  70, \dodoi{10.1088/0004-6256/143/3/70}

\bibitem[{{Amigo}(2012)}]{amigo12}
{Amigo}, P. 2012, PhD thesis, Pontificia Universidad Cat\'olica de Chile

\bibitem[{{Balona} \& {Nemec}(2012)}]{balona12}
{Balona}, L.~A., \& {Nemec}, J.~M. 2012, \mnras, 426, 2413,
  \dodoi{10.1111/j.1365-2966.2012.21957.x}

\bibitem[{{Battaglia} {et~al.}(2011){Battaglia}, {Tolstoy}, {Helmi}, {Irwin},
  {Parisi}, {Hill}, \& {Jablonka}}]{battaglia11}
{Battaglia}, G., {Tolstoy}, E., {Helmi}, A., {et~al.} 2011, \mnras, 411, 1013,
  \dodoi{10.1111/j.1365-2966.2010.17745.x}

\bibitem[{{Breger}(2000)}]{breger00}
{Breger}, M. 2000, in Astronomical Society of the Pacific Conference Series,
  Vol. 210, Delta Scuti and Related Stars, ed. M.~{Breger} \& M.~{Montgomery},
  3

\bibitem[{{Catelan} \& {Smith}(2015)}]{catelan15}
{Catelan}, M., \& {Smith}, H.~A. 2015, {Pulsating Stars}

\bibitem[{{Cicu{\'e}ndez} {et~al.}(2018){Cicu{\'e}ndez}, {Battaglia}, {Irwin},
  {Bermejo-Climent}, {McMonigal}, {Bate}, {Lewis}, {Conn}, {de Boer},
  {Gallart}, {Guglielmo}, {Ibata}, {McConnachie}, {Tolstoy}, \&
  {Fernando}}]{cicuendez18}
{Cicu{\'e}ndez}, L., {Battaglia}, G., {Irwin}, M., {et~al.} 2018, \aap, 609,
  A53, \dodoi{10.1051/0004-6361/201731450}

\bibitem[{{Clementini}(2014)}]{clementini14}
{Clementini}, G. 2014, in IAU Symposium, Vol. 301, Precision Asteroseismology,
  ed. J.~A. {Guzik}, W.~J. {Chaplin}, G.~{Handler}, \& A.~{Pigulski}, 129--136

\bibitem[{{Clementini} {et~al.}(2012){Clementini}, {Cignoni}, {Contreras
  Ramos}, {Federici}, {Ripepi}, {Marconi}, {Tosi}, \& {Musella}}]{clementini12}
{Clementini}, G., {Cignoni}, M., {Contreras Ramos}, R., {et~al.} 2012, \apj,
  756, 108, \dodoi{10.1088/0004-637X/756/2/108}

\bibitem[{{Cohen} \& {Sarajedini}(2012)}]{cohen12}
{Cohen}, R.~E., \& {Sarajedini}, A. 2012, \mnras, 419, 342,
  \dodoi{10.1111/j.1365-2966.2011.19697.x}

\bibitem[{{Coppola} {et~al.}(2015){Coppola}, {Marconi}, {Stetson}, {Bono},
  {Braga}, {Ripepi}, {Dall'Ora}, {Musella}, {Buonanno}, {Fabrizio}, {Ferraro},
  {Fiorentino}, {Iannicola}, {Monelli}, {Nonino}, {Th{\'e}venin}, \&
  {Walker}}]{coppola15}
{Coppola}, G., {Marconi}, M., {Stetson}, P.~B., {et~al.} 2015, \apj, 814, 71,
  \dodoi{10.1088/0004-637X/814/1/71}

\bibitem[{{Fiorentino} {et~al.}(2015){Fiorentino}, {Marconi}, {Bono},
  {Dalessandro}, {Ferraro}, {Lanzoni}, {Lovisi}, \&
  {Mucciarelli}}]{fiorentino15}
{Fiorentino}, G., {Marconi}, M., {Bono}, G., {et~al.} 2015, \apj, 810, 15,
  \dodoi{10.1088/0004-637X/810/1/15}

\bibitem[{{Fiorentino} \& {Monelli}(2012)}]{fiorentino12}
{Fiorentino}, G., \& {Monelli}, M. 2012, \aap, 540, A102,
  \dodoi{10.1051/0004-6361/201118621}

\bibitem[{{Flaugher} {et~al.}(2015){Flaugher}, {Diehl}, {Honscheid}, {Abbott},
  {Alvarez}, {Angstadt}, {Annis}, {Antonik}, {Ballester}, {Beaufore},
  {Bernstein}, {Bernstein}, {Bigelow}, {Bonati}, {Boprie}, {Brooks},
  {Buckley-Geer}, {Campa}, {Cardiel-Sas}, {Castander}, {Castilla}, {Cease},
  {Cela-Ruiz}, {Chappa}, {Chi}, {Cooper}, {da Costa}, {Dede}, {Derylo},
  {DePoy}, {de Vicente}, {Doel}, {Drlica-Wagner}, {Eiting}, {Elliott}, {Emes},
  {Estrada}, {Fausti Neto}, {Finley}, {Flores}, {Frieman}, {Gerdes},
  {Gladders}, {Gregory}, {Gutierrez}, {Hao}, {Holland}, {Holm}, {Huffman},
  {Jackson}, {James}, {Jonas}, {Karcher}, {Karliner}, {Kent}, {Kessler},
  {Kozlovsky}, {Kron}, {Kubik}, {Kuehn}, {Kuhlmann}, {Kuk}, {Lahav}, {Lathrop},
  {Lee}, {Levi}, {Lewis}, {Li}, {Mandrichenko}, {Marshall}, {Martinez},
  {Merritt}, {Miquel}, {Mu{\~n}oz}, {Neilsen}, {Nichol}, {Nord}, {Ogando},
  {Olsen}, {Palaio}, {Patton}, {Peoples}, {Plazas}, {Rauch}, {Reil}, {Rheault},
  {Roe}, {Rogers}, {Roodman}, {Sanchez}, {Scarpine}, {Schindler}, {Schmidt},
  {Schmitt}, {Schubnell}, {Schultz}, {Schurter}, {Scott}, {Serrano}, {Shaw},
  {Smith}, {Soares-Santos}, {Stefanik}, {Stuermer}, {Suchyta}, {Sypniewski},
  {Tarle}, {Thaler}, {Tighe}, {Tran}, {Tucker}, {Walker}, {Wang}, {Watson},
  {Weaverdyck}, {Wester}, {Woods}, {Yanny}, \& {DES
  Collaboration}}]{flaugher15}
{Flaugher}, B., {Diehl}, H.~T., {Honscheid}, K., {et~al.} 2015, \aj, 150, 150,
  \dodoi{10.1088/0004-6256/150/5/150}

\bibitem[{{F{\"o}rster} {et~al.}(2016){F{\"o}rster}, {Maureira}, {San
  Mart{\'{\i}}n}, {Hamuy}, {Mart{\'{\i}}nez}, {Huijse}, {Cabrera}, {Galbany},
  {de Jaeger}, {Gonz{\'a}lez-Gait{\'a}n}, {Anderson}, {Kunkarayakti},
  {Pignata}, {Bufano}, {Litt{\'{\i}}n}, {Olivares}, {Medina}, {Smith}, {Vivas},
  {Est{\'e}vez}, {Mu{\~n}oz}, \& {Vera}}]{forster16}
{F{\"o}rster}, F., {Maureira}, J.~C., {San Mart{\'{\i}}n}, J., {et~al.} 2016,
  \apj, 832, 155, \dodoi{10.3847/0004-637X/832/2/155}

\bibitem[{{Gaia Collaboration} {et~al.}(2018{\natexlab{a}}){Gaia
  Collaboration}, {Brown}, {Vallenari}, {Prusti}, {de Bruijne}, {Babusiaux}, \&
  {Bailer-Jones}}]{gaia18}
{Gaia Collaboration}, {Brown}, A.~G.~A., {Vallenari}, A., {et~al.}
  2018{\natexlab{a}}, ArXiv e-prints.
\newblock \doarXiv{1804.09365}

\bibitem[{{Gaia Collaboration} {et~al.}(2016){Gaia Collaboration}, {Prusti},
  {de Bruijne}, {Brown}, {Vallenari}, {Babusiaux}, {Bailer-Jones}, {Bastian},
  {Biermann}, {Evans}, \& et~al.}]{gaia16}
{Gaia Collaboration}, {Prusti}, T., {de Bruijne}, J.~H.~J., {et~al.} 2016,
  \aap, 595, A1, \dodoi{10.1051/0004-6361/201629272}

\bibitem[{{Gaia Collaboration} {et~al.}(2018{\natexlab{b}}){Gaia
  Collaboration}, {Helmi}, {van Leeuwen}, {McMillan}, {Massari}, {Antoja},
  {Robin}, {Lindegren}, {Bastian}, \& {co-authors}}]{helmi18}
{Gaia Collaboration}, {Helmi}, A., {van Leeuwen}, F., {et~al.}
  2018{\natexlab{b}}, ArXiv e-prints.
\newblock \doarXiv{1804.09381}

\bibitem[{{Garg} {et~al.}(2010){Garg}, {Cook}, {Nikolaev}, {Huber}, {Rest},
  {Becker}, {Challis}, {Clocchiatti}, {Miknaitis}, {Minniti}, {Morelli},
  {Olsen}, {Prieto}, {Suntzeff}, {Welch}, \& {Wood-Vasey}}]{garg10}
{Garg}, A., {Cook}, K.~H., {Nikolaev}, S., {et~al.} 2010, \aj, 140, 328,
  \dodoi{10.1088/0004-6256/140/2/328}

\bibitem[{{Garling} {et~al.}(2018){Garling}, {Willman}, {Sand}, {Hargis},
  {Crnojevi{\'c}}, {Bechtol}, {Carlin}, {Strader}, {Zou}, {Zhou}, {Nie},
  {Zhang}, {Zhou}, \& {Peng}}]{garling18}
{Garling}, C., {Willman}, B., {Sand}, D.~J., {et~al.} 2018, \apj, 852, 44,
  \dodoi{10.3847/1538-4357/aa9bf1}

\bibitem[{{Gautschy} \& {Saio}(2017)}]{gautschy17}
{Gautschy}, A., \& {Saio}, H. 2017, \mnras, 468, 4419,
  \dodoi{10.1093/mnras/stx811}

\bibitem[{{Irwin} \& {Hatzidimitriou}(1995)}]{irwin95}
{Irwin}, M., \& {Hatzidimitriou}, D. 1995, \mnras, 277, 1354,
  \dodoi{10.1093/mnras/277.4.1354}

\bibitem[{{Irwin} {et~al.}(1990){Irwin}, {Bunclark}, {Bridgeland}, \&
  {McMahon}}]{irwin90}
{Irwin}, M.~J., {Bunclark}, P.~S., {Bridgeland}, M.~T., \& {McMahon}, R.~G.
  1990, \mnras, 244, 16P

\bibitem[{{Kirby} {et~al.}(2011){Kirby}, {Martin}, \& {Finlator}}]{kirby11}
{Kirby}, E.~N., {Martin}, C.~L., \& {Finlator}, K. 2011, \apjl, 742, L25,
  \dodoi{10.1088/2041-8205/742/2/L25}

\bibitem[{{Lafler} \& {Kinman}(1965)}]{lafler65}
{Lafler}, J., \& {Kinman}, T.~D. 1965, \apjs, 11, 216, \dodoi{10.1086/190116}

\bibitem[{{Lee} {et~al.}(2009){Lee}, {Yuk}, {Park}, {Harris}, \&
  {Zaritsky}}]{lee09}
{Lee}, M.~G., {Yuk}, I.-S., {Park}, H.~S., {Harris}, J., \& {Zaritsky}, D.
  2009, \apj, 703, 692, \dodoi{10.1088/0004-637X/703/1/692}

\bibitem[{{Lee} {et~al.}(2003){Lee}, {Park}, {Park}, {Sohn}, {Oh}, {Yuk},
  {Rey}, {Lee}, {Lee}, {Kim}, {Han}, {Park}, {Lee}, {Jeon}, \& {Kim}}]{lee03}
{Lee}, M.~G., {Park}, H.~S., {Park}, J.-H., {et~al.} 2003, \aj, 126, 2840,
  \dodoi{10.1086/379171}

\bibitem[{{Magnier} {et~al.}(2016){Magnier}, {Schlafly}, {Finkbeiner}, {Tonry},
  {Goldman}, {R{\"o}ser}, {Schilbach}, {Chambers}, {Flewelling}, {Huber},
  {Price}, {Sweeney}, {Waters}, {Denneau}, {Draper}, {Hodapp}, {Jedicke},
  {Kudritzki}, {Metcalfe}, {Stubbs}, \& {Wainscoast}}]{magnier16}
{Magnier}, E.~A., {Schlafly}, E.~F., {Finkbeiner}, D.~P., {et~al.} 2016, ArXiv
  e-prints.
\newblock \doarXiv{1612.05242}

\bibitem[{{Maintz} \& {de Boer}(2005)}]{maintz05}
{Maintz}, G., \& {de Boer}, K.~S. 2005, \aap, 442, 229,
  \dodoi{10.1051/0004-6361:20053231}

\bibitem[{{Mart{\'{\i}}nez-V{\'a}zquez}
  {et~al.}(2016){Mart{\'{\i}}nez-V{\'a}zquez}, {Stetson}, {Monelli}, {Bernard},
  {Fiorentino}, {Gallart}, {Bono}, {Cassisi}, {Dall'Ora}, {Ferraro},
  {Iannicola}, \& {Walker}}]{martinez16}
{Mart{\'{\i}}nez-V{\'a}zquez}, C.~E., {Stetson}, P.~B., {Monelli}, M., {et~al.}
  2016, \mnras, 462, 4349, \dodoi{10.1093/mnras/stw1895}

\bibitem[{{Mateo}(1993)}]{mateo93}
{Mateo}, M. 1993, in Astronomical Society of the Pacific Conference Series,
  Vol.~53, Blue Stragglers, ed. R.~A. {Saffer}, 74

\bibitem[{{Mateo} {et~al.}(1995){Mateo}, {Fischer}, \& {Krzeminski}}]{mateo95}
{Mateo}, M., {Fischer}, P., \& {Krzeminski}, W. 1995, \aj, 110, 2166,
  \dodoi{10.1086/117676}

\bibitem[{{Mateo} {et~al.}(1996){Mateo}, {Mirabal}, {Udalski}, {Szymanski},
  {Kaluzny}, {Kubiak}, {Krzeminski}, \& {Stanek}}]{mateo96}
{Mateo}, M., {Mirabal}, N., {Udalski}, A., {et~al.} 1996, \apjl, 458, L13,
  \dodoi{10.1086/309919}

\bibitem[{{Mateo} {et~al.}(1991){Mateo}, {Nemec}, {Irwin}, \&
  {McMahon}}]{mateo91}
{Mateo}, M., {Nemec}, J., {Irwin}, M., \& {McMahon}, R. 1991, \aj, 101, 892,
  \dodoi{10.1086/115734}

\bibitem[{{McNamara}(2011)}]{mcnamara11}
{McNamara}, D.~H. 2011, \aj, 142, 110, \dodoi{10.1088/0004-6256/142/4/110}

\bibitem[{{Medina} {et~al.}(2018){Medina}, {Mu{\~n}oz}, {Vivas}, {Carlin},
  {F{\"o}rster}, {Mart{\'{\i}}nez}, {Galbany}, {Gonz{\'a}lez-Gait{\'a}n},
  {Hamuy}, {de Jaeger}, {Maureira}, \& {San Mart{\'{\i}}n}}]{medina18}
{Medina}, G.~E., {Mu{\~n}oz}, R.~R., {Vivas}, A.~K., {et~al.} 2018, \apj, 855,
  43, \dodoi{10.3847/1538-4357/aaad02}

\bibitem[{{Moretti} {et~al.}(2009){Moretti}, {Dall'Ora}, {Ripepi},
  {Clementini}, {Di Fabrizio}, {Smith}, {DeLee}, {Kuehn}, {Catelan}, {Marconi},
  {Musella}, {Beers}, \& {Kinemuchi}}]{moretti09}
{Moretti}, M.~I., {Dall'Ora}, M., {Ripepi}, V., {et~al.} 2009, \apjl, 699,
  L125, \dodoi{10.1088/0004-637X/699/2/L125}

\bibitem[{{Musella} {et~al.}(2009){Musella}, {Ripepi}, {Clementini},
  {Dall'Ora}, {Kinemuchi}, {di Fabrizio}, {Greco}, {Marconi}, {Smith},
  {Radovich}, \& {Beers}}]{musella09}
{Musella}, I., {Ripepi}, V., {Clementini}, G., {et~al.} 2009, \apjl, 695, L83,
  \dodoi{10.1088/0004-637X/695/1/L83}

\bibitem[{{Nemec} {et~al.}(1994){Nemec}, {Nemec}, \& {Lutz}}]{nemec94}
{Nemec}, J.~M., {Nemec}, A.~F.~L., \& {Lutz}, T.~E. 1994, \aj, 108, 222,
  \dodoi{10.1086/117062}

\bibitem[{{Okamoto} {et~al.}(2017){Okamoto}, {Arimoto}, {Tolstoy}, {Jablonka},
  {Irwin}, {Komiyama}, {Yamada}, \& {Onodera}}]{okamoto17}
{Okamoto}, S., {Arimoto}, N., {Tolstoy}, E., {et~al.} 2017, \mnras, 467, 208,
  \dodoi{10.1093/mnras/stx086}

\bibitem[{{Poleski} {et~al.}(2010){Poleski}, {Soszy{\'n}ski}, {Udalski},
  {Szyma{\'n}ski}, {Kubiak}, {Pietrzy{\'n}ski}, {Wyrzykowski}, {Szewczyk}, \&
  {Ulaczyk}}]{poleski10}
{Poleski}, R., {Soszy{\'n}ski}, I., {Udalski}, A., {et~al.} 2010, \actaa, 60,
  1.
\newblock \doarXiv{1004.0950}

\bibitem[{{Poretti} {et~al.}(2008){Poretti}, {Clementini}, {Held}, {Greco},
  {Mateo}, {Dell'Arciprete}, {Rizzi}, {Gullieuszik}, \& {Maio}}]{poretti08}
{Poretti}, E., {Clementini}, G., {Held}, E.~V., {et~al.} 2008, \apj, 685, 947,
  \dodoi{10.1086/591241}

\bibitem[{{Price-Whelan} {et~al.}(2018){Price-Whelan}, {Nidever}, {Choi},
  {Schlafly}, {Morton}, {Koposov}, \& {Belokurov}}]{price-whelan18}
{Price-Whelan}, A.~M., {Nidever}, D.~L., {Choi}, Y., {et~al.} 2018, ArXiv
  e-prints.
\newblock \doarXiv{1811.05991}

\bibitem[{{Pych} {et~al.}(2001){Pych}, {Kaluzny}, {Krzeminski},
  {Schwarzenberg-Czerny}, \& {Thompson}}]{pych01}
{Pych}, W., {Kaluzny}, J., {Krzeminski}, W., {Schwarzenberg-Czerny}, A., \&
  {Thompson}, I.~B. 2001, \aap, 367, 148, \dodoi{10.1051/0004-6361:20000349}

\bibitem[{{Roderick} {et~al.}(2016){Roderick}, {Jerjen}, {Da Costa}, \&
  {Mackey}}]{roderick16}
{Roderick}, T.~A., {Jerjen}, H., {Da Costa}, G.~S., \& {Mackey}, A.~D. 2016,
  \mnras, 460, 30, \dodoi{10.1093/mnras/stw949}

\bibitem[{{Saha} \& {Hoessel}(1990)}]{saha90}
{Saha}, A., \& {Hoessel}, J.~G. 1990, \aj, 99, 97, \dodoi{10.1086/115316}

\bibitem[{{Samus} {et~al.}(2017){Samus}, {Kazarovets}, {Durlevich}, {Kireeva},
  \& {Pastukhova}}]{samus17}
{Samus}, N.~N., {Kazarovets}, E.~V., {Durlevich}, O.~V., {Kireeva}, N.~N., \&
  {Pastukhova}, E.~N. 2017, Astronomy Reports, 61, 80,
  \dodoi{10.1134/S1063772917010085}

\bibitem[{{Sandage} \& {Tammann}(2006)}]{sandage06}
{Sandage}, A., \& {Tammann}, G.~A. 2006, \araa, 44, 93,
  \dodoi{10.1146/annurev.astro.43.072103.150612}

\bibitem[{{Schechter} {et~al.}(1993){Schechter}, {Mateo}, \&
  {Saha}}]{schecter93}
{Schechter}, P.~L., {Mateo}, M., \& {Saha}, A. 1993, \pasp, 105, 1342,
  \dodoi{10.1086/133316}

\bibitem[{{Schlafly} \& {Finkbeiner}(2011)}]{schlafly11}
{Schlafly}, E.~F., \& {Finkbeiner}, D.~P. 2011, \apj, 737, 103,
  \dodoi{10.1088/0004-637X/737/2/103}

\bibitem[{{Schlegel} {et~al.}(1998){Schlegel}, {Finkbeiner}, \&
  {Davis}}]{schlegel98}
{Schlegel}, D.~J., {Finkbeiner}, D.~P., \& {Davis}, M. 1998, \apj, 500, 525,
  \dodoi{10.1086/305772}

\bibitem[{{Sesar} {et~al.}(2010){Sesar}, {Ivezi{\'c}}, {Grammer}, {Morgan},
  {Becker}, {Juri{\'c}}, {De Lee}, {Annis}, {Beers}, {Fan}, {Lupton}, {Gunn},
  {Knapp}, {Jiang}, {Jester}, {Johnston}, \& {Lampeitl}}]{sesar10}
{Sesar}, B., {Ivezi{\'c}}, {\v Z}., {Grammer}, S.~H., {et~al.} 2010, \apj, 708,
  717, \dodoi{10.1088/0004-637X/708/1/717}

\bibitem[{{Sesar} {et~al.}(2017){Sesar}, {Hernitschek}, {Mitrovi{\'c}},
  {Ivezi{\'c}}, {Rix}, {Cohen}, {Bernard}, {Grebel}, {Martin}, {Schlafly},
  {Burgett}, {Draper}, {Flewelling}, {Kaiser}, {Kudritzki}, {Magnier},
  {Metcalfe}, {Tonry}, \& {Waters}}]{sesar17}
{Sesar}, B., {Hernitschek}, N., {Mitrovi{\'c}}, S., {et~al.} 2017, \aj, 153,
  204, \dodoi{10.3847/1538-3881/aa661b}

\bibitem[{{Skarka}(2014)}]{skarka14}
{Skarka}, M. 2014, \mnras, 445, 1584, \dodoi{10.1093/mnras/stu1815}

\bibitem[{{Stetson} {et~al.}(2014){Stetson}, {Fiorentino}, {Bono}, {Bernard},
  {Monelli}, {Iannicola}, {Gallart}, \& {Ferraro}}]{stetson14}
{Stetson}, P.~B., {Fiorentino}, G., {Bono}, G., {et~al.} 2014, \pasp, 126, 616,
  \dodoi{10.1086/677352}

\bibitem[{{Taylor}(2006)}]{taylor06}
{Taylor}, M.~B. 2006, in Astronomical Society of the Pacific Conference Series,
  Vol. 351, Astronomical Data Analysis Software and Systems XV, ed.
  C.~{Gabriel}, C.~{Arviset}, D.~{Ponz}, \& S.~{Enrique}, 666

\bibitem[{{Tonry} {et~al.}(2012){Tonry}, {Stubbs}, {Lykke}, {Doherty},
  {Shivvers}, {Burgett}, {Chambers}, {Hodapp}, {Kaiser}, {Kudritzki},
  {Magnier}, {Morgan}, {Price}, \& {Wainscoat}}]{tonry12}
{Tonry}, J.~L., {Stubbs}, C.~W., {Lykke}, K.~R., {et~al.} 2012, \apj, 750, 99,
  \dodoi{10.1088/0004-637X/750/2/99}

\bibitem[{{Valdes} {et~al.}(2014){Valdes}, {Gruendl}, \& {DES
  Project}}]{valdes14}
{Valdes}, F., {Gruendl}, R., \& {DES Project}. 2014, in Astronomical Society of
  the Pacific Conference Series, Vol. 485, Astronomical Data Analysis Software
  and Systems XXIII, ed. N.~{Manset} \& P.~{Forshay}, 379

\bibitem[{{Vivas} \& {Mateo}(2013)}]{vivas13}
{Vivas}, A.~K., \& {Mateo}, M. 2013, \aj, 146, 141,
  \dodoi{10.1088/0004-6256/146/6/141}

\bibitem[{{Vivas} {et~al.}(2001){Vivas}, {Zinn}, {Andrews}, {Bailyn}, {Baltay},
  {Coppi}, {Ellman}, {Girard}, {Rabinowitz}, {Schaefer}, {Shin}, {Snyder},
  {Sofia}, {van Altena}, {Abad}, {Bongiovanni}, {Brice{\~n}o}, {Bruzual},
  {Della Prugna}, {Herrera}, {Magris}, {Mateu}, {Pacheco}, {S{\'a}nchez},
  {S{\'a}nchez}, {Schenner}, {Stock}, {Vicente}, {Vieira}, {Ferr{\'{\i}}n},
  {Hernandez}, {Gebhard}, {Honeycutt}, {Mufson}, {Musser}, \&
  {Rengstorf}}]{vivas01}
{Vivas}, A.~K., {Zinn}, R., {Andrews}, P., {et~al.} 2001, \apjl, 554, L33,
  \dodoi{10.1086/320915}

\bibitem[{{Zinn} {et~al.}(2014){Zinn}, {Horowitz}, {Vivas}, {Baltay}, {Ellman},
  {Hadjiyska}, {Rabinowitz}, \& {Miller}}]{zinn14}
{Zinn}, R., {Horowitz}, B., {Vivas}, A.~K., {et~al.} 2014, \apj, 781, 22,
  \dodoi{10.1088/0004-637X/781/1/22}

\end{thebibliography}


\end{document}